\documentclass{article}
\usepackage{mrs2005,epsfig}
\begin{document} 
\title{CHEMICAL ELEMENTS AT HIGH AND LOW REDSHIFTS}

\author{MAX PETTINI}
\affil{Institute of Astronomy, Cambridge, CB3 0HA, England}

\begin{abstract} 
The past few years have seen a steady progress in the determination
of element abundances at high redshifts, with new and more accurate
measures of metallicities in star-forming galaxies, in
QSO absorbers, and in the intergalactic medium. We have also
become more aware of the limitations of the tools at our disposal
in such endeavours. I summarise these
recent developments and---in tune with 
the theme of this meeting---consider
the clues which chemical abundance studies offer 
to the links between the high redshift galaxy populations 
and today's galaxies. The new data are `fleshing out'
the overall picture of element abundances at redshifts 
$z = 2 - 3$ which has been gradually coming into focus
over the last decade. In particular, we can now account for at least
40\% of the metals produced by the global star formation activity
in the universe from the Big Bang to $z = 2.5$, and we have strong
indications of where the remainder are likely to be found.
\end{abstract} 
 
\section{Introduction} 
 
Studies of element abundances, like many other areas of astrophysics,
have undergone a remarkable acceleration in the flow of data over 
the last few years. 
We have witnessed wholesale abundance determinations in tens of 
thousands of galaxies from large scale surveys such as the 2dF
galaxy redshift survey and the Sloan Digital Sky Survey (SDSS),
measurements of the abundances of different elements in individual
stars of Local Group galaxies beyond the immediate vicinity of the 
Milky Way, and the determination with exquisite precision of the 
detailed chemical composition of some of the first stars to form in
the Galactic halo, with less than one thousandth of the heavy elements
which were present when the Sun formed. 
Chemical abundance studies are also increasingly being extended
to high redshifts, charting the progress of stellar nucleosynthesis
over most of the age of the universe.
These are all projects which
were well beyond our capabilities only a decade ago.

The main motivation common to all of these observational efforts 
is to use the chemical information as one of the means at our disposal
to link the properties of high redshift galaxies with those we see 
around us today and thereby understand the physical processes at 
play in the formation and the evolution of galaxies. This is indeed the 
theme which has inspired this meeting. Viewed from this perspective,
the remarkable observational advances of the last few years
have yet `to deliver' fully. That is,
we are still struggling to understand and fit together into
a coherent picture of the chemical evolution of galaxies
all this new information which we have been gathering at 
such a fast pace. It is this effort which will be the main focus
of my review, which I have divided into four parts.

First I shall review the tools at our disposal for measuring
element abundances, which
themselves are far from being firmly established and calibrated,
but rather have been undergoing significant scrutiny and
revisions of late.
Second, I shall present recent results on  
element abundances at high redshift in star-forming galaxies 
and in QSO absorption line systems.
I shall then consider the clues which these measurements
offer us in our quest to establish a connection
between these high redshift populations and 
their descendants in today's universe.
The review concludes with an overall assessment of our
knowledge of element abundances in different components
of the universe at $z = 2.5$ and their evolution to the present time.

\section{Abundance Measurement Techniques: Recent Developments} 

\subsection{Emission Line Abundances}

Nebular emission lines from H~{\sc ii} regions have been the main
tool at our disposal for measuring element abundances, and 
their radial gradients, in low redshift galaxies. In general,
one has to rely on some empirical calibration of the strongest
emission lines which are the ones most easily observed.
Much work has been devoted to these calibrations in recent
years, but there is still significant dispersion in the metallicities
implied by different indices and, more worryingly, between
different calibrations of the same index.

Recently, using large aperture telescopes, it has become possible
to detect weak, temperature sensitive, 
auroral lines in extragalactic metal-rich H~{\sc ii} regions
(e.g. Garnett, Bresolin, \& Kennicutt 2004).
These lines should, at least in
principle, allow an accurate measure of the temperature
of the H~{\sc ii} regions studied and thereby put the 
abundance determinations on a more solid footing. 
The results so far
(several groups are now involved in this type of work---see
also Sara Ellison's contribution to these proceedings) 
suggest that the well-used $R23$ index of Pagel et al. (1979)
may overestimate the oxygen abundance 
significantly at high metallicities---and in case you should
think that high metallicities are not relevant at high redshifts,
think again!
There is still a degree of controversy as to what extent
these temperature-based abundances should be trusted
(e.g. Stasi{\' n}ska 2005).
If confirmed by future work, the implications of these
initial studies are profound: super-solar metallicities 
are much rarer than previously envisaged (which may 
be in better accord with the predictions of chemical 
evolution models and with the paucity of super-metal-rich 
stars in the inner regions of the Milky Way) and galactic 
abundance gradients are more shallow than we were led to 
believe. 

When it comes to applying these line diagnostics at high 
redshifts we are faced with additional problems, apart
from the obvious one of a reduction in the flux received.
When the emission lines are redshifted to near-infrared 
wavelengths
only a subset of the strong lines is normally accessible, free
from water vapour absorption and OH emission from the
night sky. Even at the highly convenient redshift of $z \simeq 2.3$, 
where all the strong lines fall into near-IR windows, there is 
at present only
one near-IR spectrograph on a large telescope, the Gemini
GNIRS, which allows one to record all the strong lines,
from [O~{\sc ii}]~$\lambda 3727$ to H$\alpha$ 
and [S~{\sc ii}]~$\lambda \lambda 6716, 6731$, 
simultaneously.
For this reason, an abundance index based on the ratio of
only two lines which are close in wavelength is particularly 
attractive. Despite concerns about its dependence on the 
ionisation parameter, the $N2$ index (see Figure 1)
recently recalibrated by
Pettini \& Pagel (2004) has so far proved to be the most direct 
way to obtain at least an approximate estimate of the 
degree of metal enrichment of star-forming galaxies at
$z = 1.5 - 2.5$. \\

%
%
\begin{figure}  
\begin{center}
{\hspace*{-0.5cm} \epsfig{figure=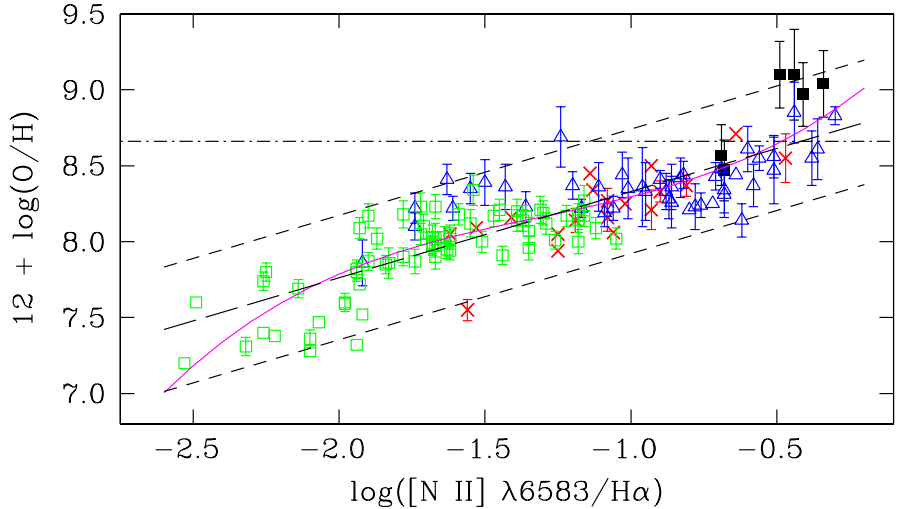,width=7.5cm,angle=270}}  
\end{center}
\vspace*{-0.5cm}  
\caption{The empirical calibration by Pettini \& Pagel (2004) of the 
$N2$ index, defined as $\log\,({\rm [N~II]~\lambda 6583/H\alpha}$).
Symbols are for individual H~{\sc ii} regions with measured electron
temperature or where the oxygen abundance has been derived with
detailed photoionisation models---see Pettini \& Pagel (2004) for references
to the original works. 
The long-dash line is the best fitting linear relationship:
$12 + \log {\rm (O/H)} = 8.90 + 0.57 \times N2$. 
The short-dash lines encompass 95\% of the measurements
and correspond to a range in $\log {\rm (O/H)} = \pm 0.41$
relative to this linear fit.
The continuous line is a cubic function of the form
$12 + \log {\rm (O/H)} = 
9.37 + 2.03 \times N2 + 1.26 \times N2^{2} + 0.32 \times N2^{3}$
which, however, gives only a slightly better fit to the data
(95\% of the data points are within $\pm 0.38$ of this line).
The dot-dash horizontal line shows the solar oxygen abundance
$12 + \log {\rm (O/H)} = 8.66$ (Asplund et al. 2004). 
} 
\end{figure} 

\subsection{Absorption Line Abundances}

Resonance absorption lines in the rest frame ultraviolet are the main
tool at our disposal for determining the abundances of a wide
variety of chemical elements in cool interstellar gas. They have
been applied primarily to the interstellar medium (ISM)
of the Milky Way and Magellanic
Clouds, and to high redshift absorption systems seen in the
spectra of QSOs. This technique is capable of achieving high
precision, provided one is in a regime where the absorption
line optical depth is low (i.e. line saturation is not a concern),
ionisation corrections are negligible, and dust depletions can be 
accounted for. These are all tractable problems in damped 
Lyman~$\alpha$ systems (DLAs), the class of QSO absorbers with
the largest column densities of neutral gas (Wolfe, Prochaska, \& Gawiser 2005).
However, this is generally \emph{not} the case in star-forming galaxies
at high $z$, where normally: (i) only the
strongest (and therefore saturated) lines are measured; 
(ii) $N$(H~{\sc i}) is
unavailable (given that the Ly$\alpha$ line is a complex 
blend of emission and absorption); 
(iii)  the spectrum is a composite
of multiple sightlines, so that there may be issues of non-uniform coverage
of the background source against which the absorption lines are 
detected; and (d) the ionisation structure of the ISM 
is likely to be far from simple.\\

\subsection{Stellar Abundance Diagnostics} 
The UV spectra of star-forming galaxies are rich in stellar 
features, including photospheric lines and P-Cygni features produced in the 
outflowing winds of the most luminous stars. 
Given the difficulties with the interstellar absorption lines outlined above,
attention has turned to the stellar spectrum in the search for useful
abundance diagnostics (e.g. Leitherer et al. 2001). 
The sophistication of modern non-LTE, 
line blanketed, stellar atmosphere codes, such as 
\emph{WM-basic} (Pauldrach et al. 2001), and  
of stellar population spectral synthesis codes, 
such as \emph{Starburst99} (Leitherer et al. 1999), 
allowed Rix et al. (2004) to fully synthesise the emergent spectrum
of a star-forming galaxy for a variety of star formation histories and
metallicities. As expected, it is the wind lines---particularly 
Si~{\sc iv}~$\lambda\lambda 1393, 1402$ and 
C~{\sc iv}~$\lambda\lambda 1548, 1550$---which respond
most sensitively to metallicity. In practice, however, these resonance
lines are usually blended with strong interstellar components in 
the spectra of starburst galaxies, near and far, so that their measurement
is not straightforward. Rix et al. (2004) also quantified the dependence on
metallicity of a number of blends of stellar photospheric lines, 
particularly a broad feature due to Fe~{\sc iii} at wavelengths 
between 1880 and 2020\,\AA\ (see Figure 2). 
Their main conclusion
is that these features can indeed be calibrated against metallicity $Z$,
although their broad and shallow nature requires spectra of higher
signal-to-noise ratio (S/N) than that normally afforded by current instrumentation.
Thus, while at present UV stellar features provide mostly
a consistency check on other, more straightforward, metallicity
indicators, I expect them to play a more important role in chemical
abundance studies in the future, when the next generation of very large
telescopes will routinely give us good quality spectra of high redshift
galaxies.

%
%
\begin{figure}[h]
\vspace*{-3.75cm}  
\begin{center}
{\hspace*{-0.75cm} \epsfig{figure=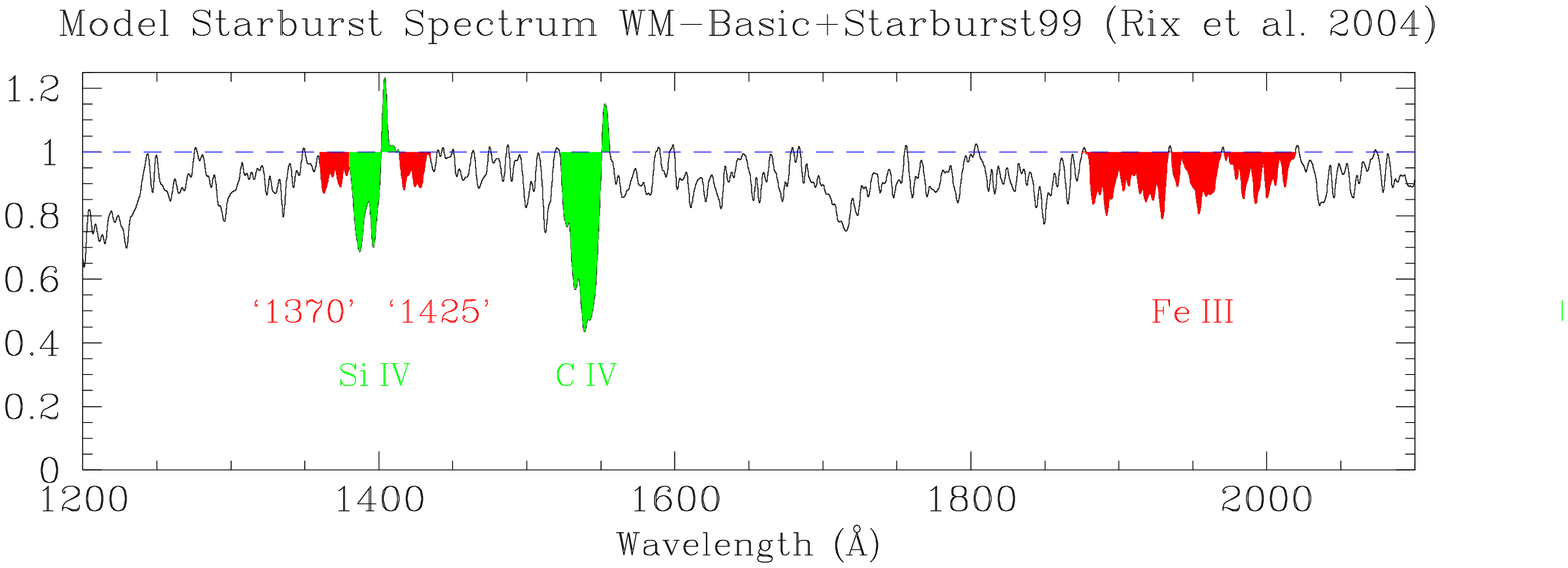,width=13.5cm,angle=270}}  
\end{center}
\vspace{-3.95cm}  
\caption{Fully theoretical UV spectrum of a starburst galaxy computed 
by Rix et al. (2004) for the case of continuous star formation with a Salpeter
initial mass function (IMF) and solar metallicity. The shaded regions indicate 
stellar wind lines (green) and blends of photospheric lines (red) 
found by Rix et al. to be sensitive to metallicity.
}
\end{figure} 

\section{Abundance Measurements: Recent Results} 

\subsection{Star-Forming Galaxies}

The largest sample of high redshift galaxies with abundance determinations
consists of nearly 100 UV-selected galaxies at redshifts $z \simeq 2.2$ satisfying
the `BX' colour criteria of Adelberger  et al. (2004), and whose
general properties have been described by Steidel et al. (2004). 
Erb et al. (2006a,b) used the near-infrared spectrograph 
NIRSPEC on the Keck~II telescope in the $K$-band
to record the spectral region encompassing the 
H$\alpha$ emission line and constructed composite spectra
of sufficiently high signal-to-noise ratio to allow the weaker [N~{\sc ii}]
(and [S~{\sc ii}]) emission lines to be measured. Application of the
$N2$ index calibration of Pettini \& Pagel (2004) then yields average 
values of the oxygen abundance for different subsets of these galaxies.
Two examples are shown in Figure 3, constructed
from galaxies respectively brighter and fainter than $K_s = 20$;
the former exhibit approximately solar (O/H) and even in the 
fainter ones the average metallicity is only a factor
of $\sim 2$ below solar (Shapley et al. 2004).
A clearer pattern emerges when bins in stellar mass are
considered:  Erb et al. (2006a) found that
galaxies which are young and have turned only a small fraction 
of their baryons into stars have (O/H)\,$ < 1/3$ solar, while in galaxies
which have already assembled $\sim 10^{11}\, M_{\odot}$ in stars
(and galaxies with $K_s < 20$ tend to be
mostly in this category) the oxygen abundance is close to solar.

%
%
\begin{figure}
\begin{center}
\vspace*{-0.5cm}
{\hspace*{-0.65cm} \epsfig{figure=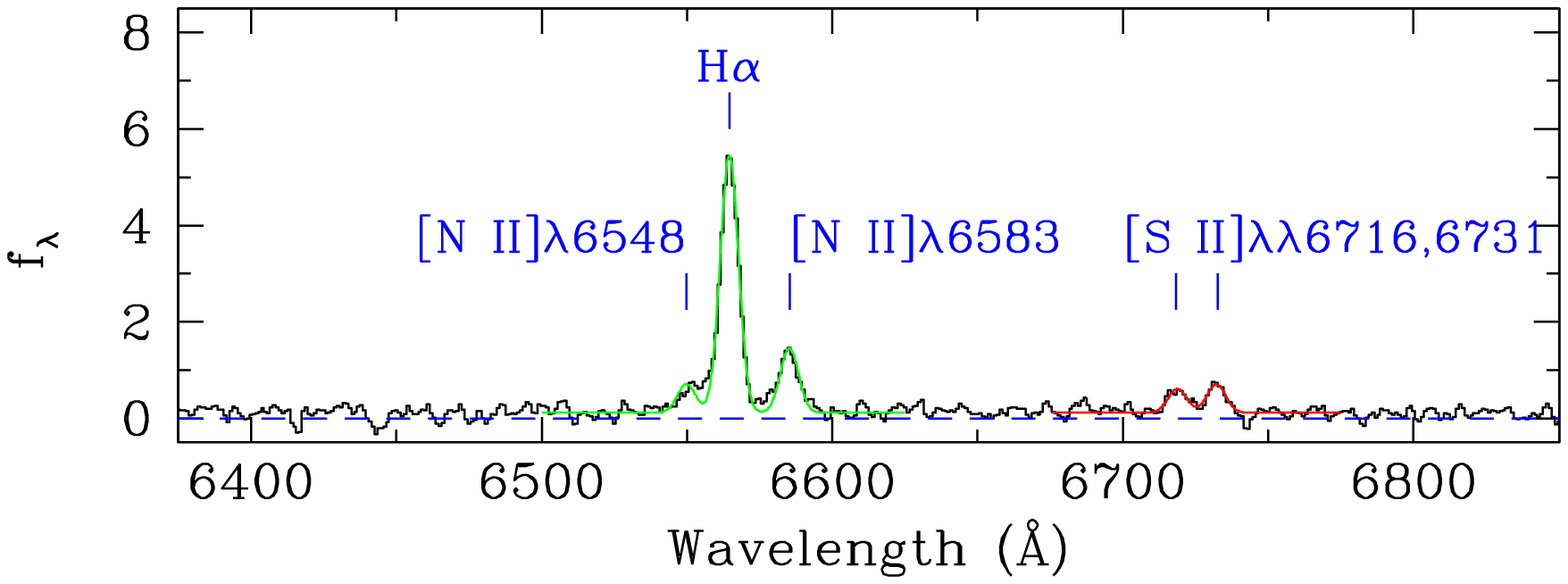,width=4.5cm,viewport=195 125 400 650,clip,angle=270}}  
{\hspace*{-0.65cm} \epsfig{figure=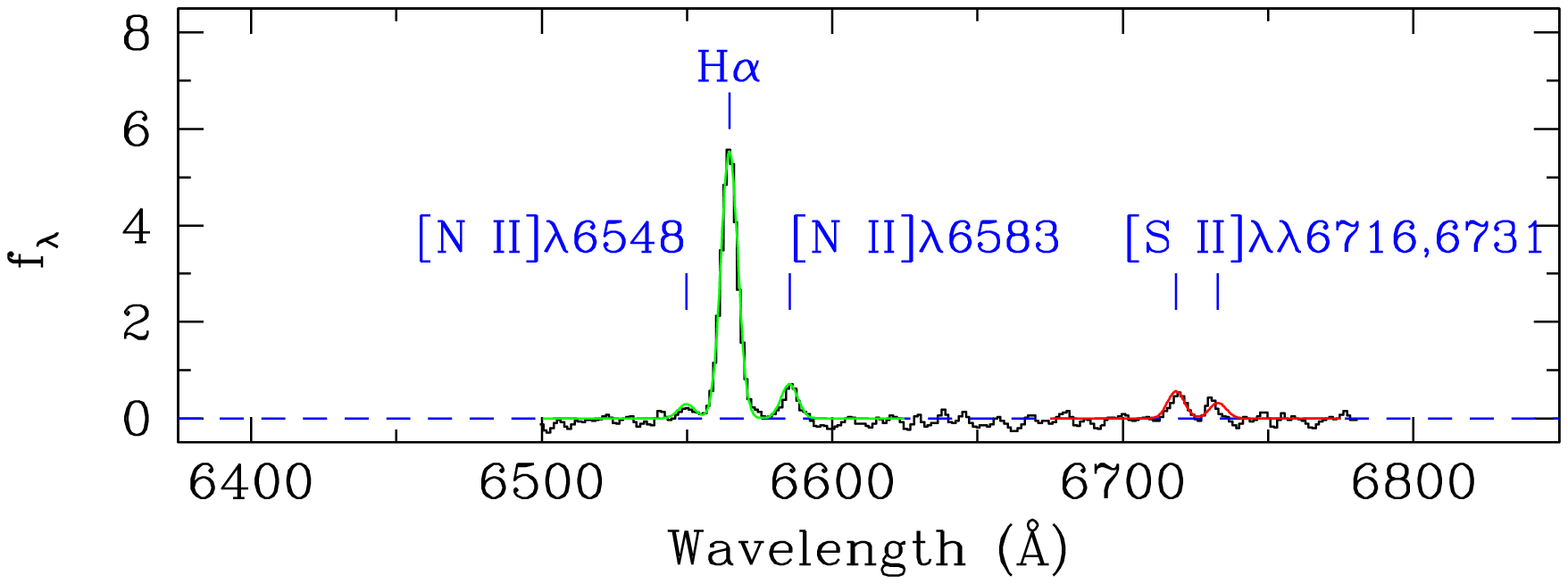,width=4.55cm,viewport=195 125 400 650,clip,angle=270}}  
\end{center}
\vspace{-0.25cm}
\caption{Composite spectra of  BX galaxies at $z \simeq 2.2$
from the survey by Erb et al. (2006b). 
\emph{Upper Panel:} galaxies brighter than $K_s = 20$ have a mean
ratio [N~{\sc ii}]/H$\alpha = 0.25$ which indicates an oxygen abundance
$12 + \log {\rm (O/H)} = 8.56$, or $\sim 4/5$ solar, if the local calibration
of the $N2$ index with (O/H) applies to these galaxies. 
\emph{Lower Panel:} galaxies fainter than $K_s = 20$ have 
[N~{\sc ii}]/H$\alpha = 0.13$ and 
$12 + \log {\rm (O/H)} = 8.39$ or $\sim 1/2$ solar.
}
\end{figure} 
%
%
\begin{figure}
\begin{center}
\vspace*{-1.5cm}
{\hspace*{-1.95cm} \epsfig{figure=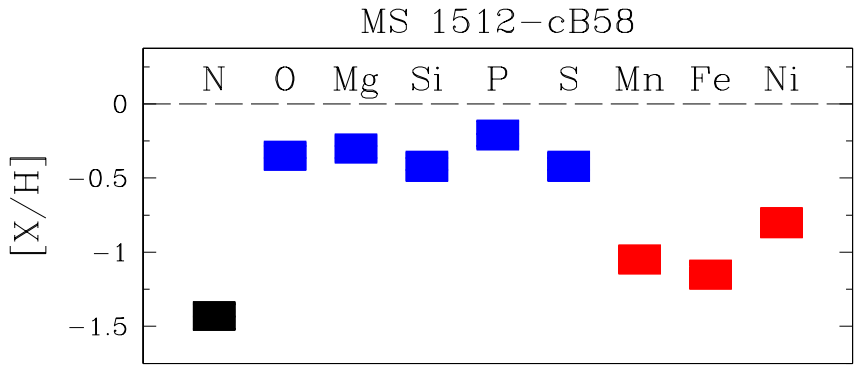,width=11cm,angle=270}}  
\end{center}
\vspace{-3.5cm}
\caption{Element abundances in the ambient interstellar medium of
the Lyman break galaxy MS~1512-cB58, at $z = 2.7276$, determined
by Pettini et al. (2002). I use the conventional shorthand whereby 
[X/H]$_{\rm cB58} = \log{\rm (X/H)}_{\rm cB58} - \log {\rm (X/H)}_{\odot}$.
While O and other $\alpha$-capture elements
(blue rectangles) have already reached $\sim 2/5$ of their solar values,
the production and mixing within the ISM of Fe-peak (red) and N (black)
apparently lag behind (even when account is taken of likely dust depletions).
If N is produced mostly by stars in the mass range $3 - 7 M_{\odot}$,
the implied timescale for the metal enrichment of the ISM is very rapid,
only a few hundred million years (Matteucci \& Pipino 2002).
MS~1512-cB58 is still the only high redshift galaxy where this type
of refined abundance analysis has been possible, thanks to gravitational
lensing by a foreground galaxy cluster which magnifies the signal 
by a factor of $\sim 25$.
}
\end{figure} 

\noindent The conclusion that many star-forming galaxies had already gained
near-solar metallicities at $z = 2-3$, while perhaps surprising to some,
has in fact been reached independently by several groups applying
different abundance diagnostics (some more reliable than others) to
galaxies selected by different techniques, including near-IR and 
sub-mm selection (e.g. Mehlert et al. 2002; Savaglio et al. 2004; 
de Mello et al. 2004; Swinbank et al. 2004). In all of these cases, 
however, we are dealing with relatively bright objects, 
more luminous than $\sim 1/4L^{\ast}$.
While the work of Erb et al. (2006a) has shown that there is no
clear luminosity-metallicity relation in these galaxies---given 
the wide spread in their mass-to-light ratios
even at rest frame near-IR wavelengths 
(Shapley et al. 2005)---it is still likely 
that galaxies with luminosities $L \ll L^{\ast}$ 
are in general less chemically enriched.

The `metallicity' deduced by all of the above studies is only
a rough measure of the overall degree of metal enrichment
of the galaxies considered. Much more information is potentially
available from their detailed chemical composition, but to date
there is only one case which has been accessible to this kind of
analysis: the gravitationally lensed galaxy MS~1512-cB58 
at $z = 2.7276$ (see Figure 4).
The relative underabundance of elements produced by intermediate 
and low mass stars, compared to the products of Type~II supernovae, 
is highly suggestive of a rapid timescale for the build-up of metals, 
of the order of a few hundred million years, consistent with the very
high star formation rates---typically tens of solar masses per 
year---exhibited by these UV-bright galaxies.\\

\subsection{Damped Lyman~$\alpha$ Systems}

Abundance measurements are now also available for
over one hundred damped Ly$\alpha$ systems. 
Zinc has
played an important role in this context because, unlike most
other Fe-peak elements,  it is
normally undepleted onto dust and the resonance
doublet of its major ionisation stage in H~{\sc i} regions,
Zn~{\sc ii}~$\lambda \lambda 2026, 2062$, 
is well placed for observations from the ground over a 
wide range of redshifts (Pettini, Boksenberg, \& Hunstead 1990).
The latest published compilation of [Zn/H] 
determinations in DLAs 
by Kulkarni et al. (2005) includes
87 measurements or upper limits, mostly obtained from echelle
spectroscopy with large telescopes (see Figure 5). 
It is clear from these data that absorption-selected galaxies give 
quite a different picture of chemical enrichment at high redshift
from that obtained when considering the bright galaxies detected 
directly via their emission. First, DLAs are mostly metal poor.
Adding up all the Zn and H separately, we obtain the column
density-weighted mean metallicity 
$[ \langle {\rm Zn/H} \rangle] = -1.2$ 
$\left ( {\rm or} ~Z \simeq 1/15 Z_{\odot} \right ) $ at $1.8 < z < 3.5$.
Second, there is a wide dispersion in metallicity 
among DLA galaxies at the same epoch, 
with individual values of the Zn abundance spanning
two orders of magnitude, from solar to less than 1/100 of
solar.\\

%
%
\begin{figure}[h]  
\begin{center}
\vspace*{-3.75cm}
{\hspace*{-1.15cm} \epsfig{figure=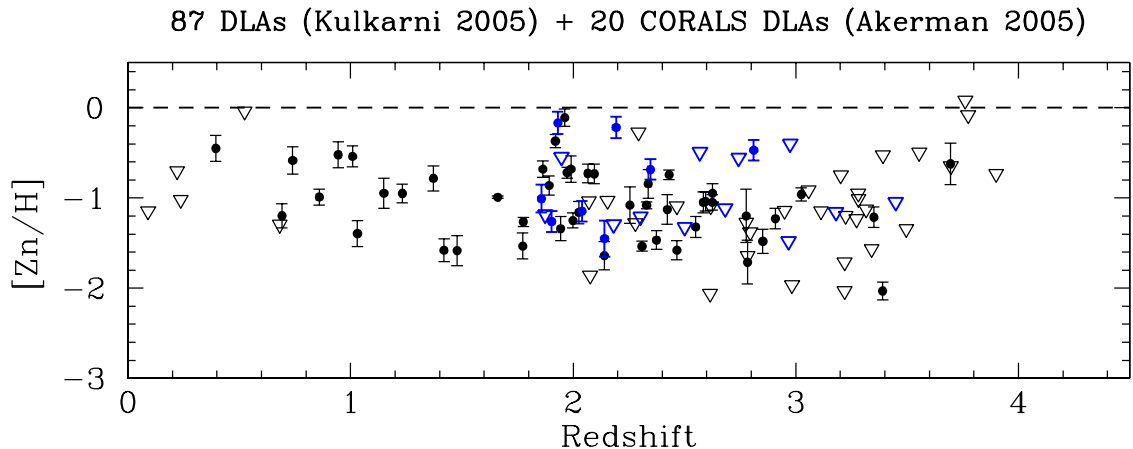,width=14.5cm,angle=270}}  
\end{center}
\vspace{-4.7cm}
\caption{Zn abundances in DLAs, from the compilation by
Kulkarni et al. 2005 (black), which brings together the results of
several surveys for DLAs in optically selected QSO samples,
and from the recent survey of CORALS 
(radio-selected) QSOs by Akerman et al. 2005 (blue). 
Triangles denote upper limits in DLAs whose 
Zn~{\sc ii}~$\lambda \lambda 2026, 2062$
lines remain undetected.
}
\end{figure} 

\subsubsection{Dust-induced bias?}

The consistently low metallicity of DLAs at all redshifts has
been a cause of concern for models which use them to track 
the global evolution of baryons through the cosmic ages
(e.g. Pei \& Fall 1995;  Kulkarni \& Fall 2002; 
Nagamine, Springel, \& Hernquist 2004;
Wolfe et al. 2005).
The biasing effects of dust have been invoked by many 
as a possible explanation of this apparent puzzle. DLAs
are generally culled from QSO surveys which are magnitude limited
and the brighter ones tend to be selected for subsequent spectroscopic
follow-up for practical reasons of S/N and resolution. 
It thus seems plausible that the most metal-rich members
of a distribution of metallicities may be preferentially missing 
from current samples, 
if the background QSOs against which they would be seen
are sufficiently dimmed by dust in the very absorbers 
being sought. 

Despite the appeal of its simplicity, the possibility that dust
may be significantly distorting our view of high redshift QSOs
and their intervening absorbers has turned out to be unsupported
by the available data: DLAs with the required
characteristics appear to be genuinely rare since they have not
been found even in surveys which use radio-selected QSOs
as background sources, where dust bias is presumably not an issue
(Ellison et al. 2001). As can be appreciated from Figure 5, 
the Zn abundance of DLAs in the (radio-selected) CORALS sample
is not significantly different than that of the bulk of known
DLAs drawn from magnitude limited optical samples of QSOs.
Akerman et al. (2005) deduced a column density-weighted metallicity
$[ \langle {\rm Zn/H} \rangle] = -0.87 \pm 0.13$ for CORALS DLAs,
only marginally higher than the corresponding value
$[ \langle {\rm Zn/H} \rangle] = -1.17 \pm 0.07$ 
determined from optical surveys over the same redshift interval, 
and concluded that the two sets of data are consistent with being 
drawn from the same parent distribution at the 90\% confidence level. 

The recent detection of dust reddening associated with 
Ca~{\sc ii}-selected DLAs at $z \sim 1$ in SDSS QSOs 
(Wild, Hewett, \& Pettini 2006) confirms that the level
of extinction remains generally low until these  
later epochs, since even this subset of absorbers,
with rather extreme properties compared to typical DLAs,
introduces only a modest $\langle E(B-V) \rangle < 0.1$\,mag.

\subsubsection{Systematic offset between absorption- and emission-measured
abundances?}

The apparent `disconnect' in metallicity between star-forming
galaxies and DLAs at the same cosmic epochs has led some to
question whether there might be systematic offsets between
abundances measured in H~{\sc ii} and H~{\sc i} regions---a
possibility apparently admitted by recent 
\emph{Far-Ultraviolet Spectroscopic Explorer} 
(FUSE) observations of low metallicity dwarf galaxies,
although the interpretation of the FUSE data is far from
straightforward (Cannon et al. 2005).
Such offsets, if present, may be due to: (a) systematic
errors in the abundance determinations; (b) abundance gradients
between the inner, high surface density, star-forming regions of a galaxy
and its outer regions which present the larger cross-section for
absorption against a background source; or (c) in a more general
sense, the existence of extended envelopes of unprocessed gas
on the outskirts of galaxies. \\

%
%
\begin{figure}[h]
\begin{center}
{\hspace*{-0.005cm} \epsfig{figure=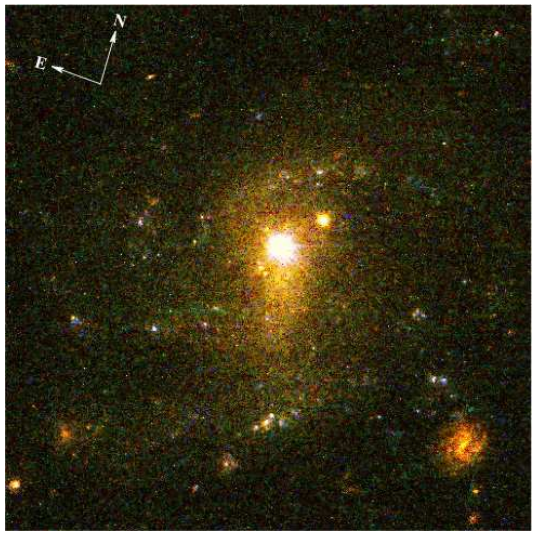,width=7cm}}  
\end{center}
\caption{\emph{Hubble Space Telescope} (WFPC2) image of t\emph{}he 
nearby ($d = 43.3 \, h_{70}^{-2}$\,Mpc) 
dwarf ($M_B = -16.8$) galaxy SBS~1543+593 
(reproduced from Schulte-Ladbeck et al. 2004).
The portion of the WFPC2 image shown here is $50 \times 50$ arcsec,
corresponding to $10.4 \times 10.4$\,kpc at the distance of the galaxy.
The bright source near the centre of the galaxy is not its nucleus,
but a background ($z_{\rm em} = 0.807$) QSO in whose spectrum
the ISM of the dwarf galaxy produces a DLA with
$\log N$(H~{\sc i})\,$ =  20.42$ (where $N$ is the column
density in units of cm$^{-2}$).
The metallicity of the DLA is the same as that of the brightest
H~{\sc ii} region in the galaxy, located 3.3\,kpc directly below
its centre in this image.
}
\end{figure} 

\noindent The `cleanest' test of these hypotheses was recently performed
by Bowen et al. (2005) who found the abundance of S in the neutral
ISM of the dwarf galaxy SBS~1543+593 to be the same as that of
O in an H~{\sc ii} region 3.3\,kpc away (see Figure 6).
This comparison is particularly meaningful because S and O
are both alpha-capture elements (i.e. they share similar nucleosynthetic
histories) and corrections for dust depletion are normally small
for both elements. This reassuring consistency check excludes options
(a) and (c) above---at least in the case of this galaxy. However, it provides 
no information on the importance of abundance gradients, particularly
at high redshifts, because today's dwarf galaxies 
do not exhibit significant radial gradients in their
metallicity.

\subsubsection{Clues from element ratios.}

As well as exhibiting widely different degrees of metal enrichment,
DLAs also appear to be quite a `mixed bag' in their detailed
chemical composition. For example, no clear pattern emerges 
from consideration of the relative proportions of alpha-capture
to Fe-peak elements. In DLAs the [$\alpha$/Fe] ratio is most
conveniently measured via [S/Zn], two elements which show little
affinity for dust and whose first ions (the dominant ion stages
in H~{\sc i} regions) have absorption lines which generally are not
heavily saturated in DLAs. On the other hand, 
the fact that these transitions are separated by 
nearly 800\,\AA\ in the rest frame ultraviolet restricts
the redshift range over which they can both be measured 
in the same absorption system. For this reason, the current
sample of [S/Zn] measurements in DLAs is still relatively
small (see Figure~7). Nevertheless, the available data
indicate that while some DLAs may conform
to the Milky Way pattern of enhanced [$\alpha$/Fe] ratios
at metallicities [Fe/H]\,$\la -1$, many of them 
evidently do not, exhibiting solar or even sub-solar 
values of [S/Zn]. 
In the context of current galactic chemical evolution models,
such low ratios are interpreted as evidence for generally low, 
or intermittent, rates of star formation
(e.g. Pagel \& Tautvaisiene 1995; Matteucci \& Recchi 2001).

%
%
\begin{figure}[h]
\begin{center}
\vspace*{-5.5cm}
{\hspace*{-1.4cm} \epsfig{figure=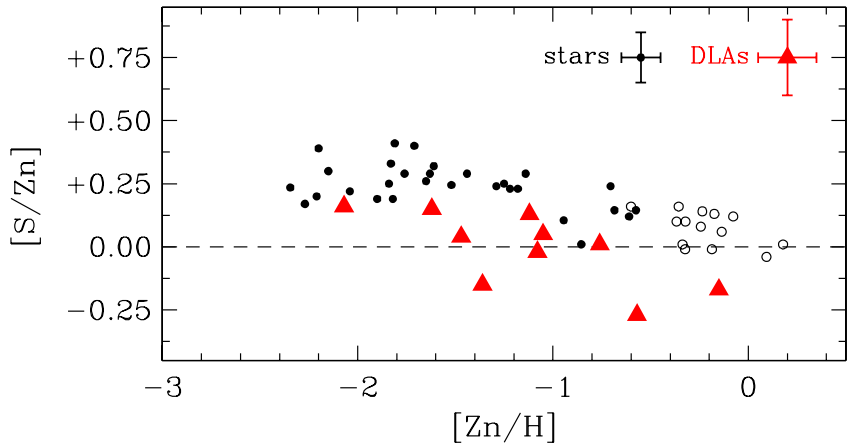,width=13.5cm}}  
\end{center}
\vspace{-6.5cm}
\caption{ Relative abundances of S (an $\alpha$-capture element)
and Zn (an Fe-peak element) in Galactic disk (open circles) and
halo (filled circles) stars and in DLAs (red triangles). 
Typical errors in the
abundance determinations are shown in the top right-hand corner.
The existing, small, sample of damped systems in which both
S and Zn have been measured does not show clear evidence
of the $\alpha$-element enhancement at low metallicities typical
of Milky Way stars, but a wider spread of [$\alpha$/Fe] ratios---a
further demonstration of the heterogeneous nature of DLA galaxies.
(Figure reproduced from Nissen et al. 2004).}
\end{figure} 

\section{The Link to Today's Galaxies}

\subsection {Star-Forming Galaxies at $z \ga 2$:  Ellipticals
and Bulges in the Making}

%
%
\begin{figure}[h] 
\begin{center}
{\hspace*{-0.3cm} \epsfig{figure=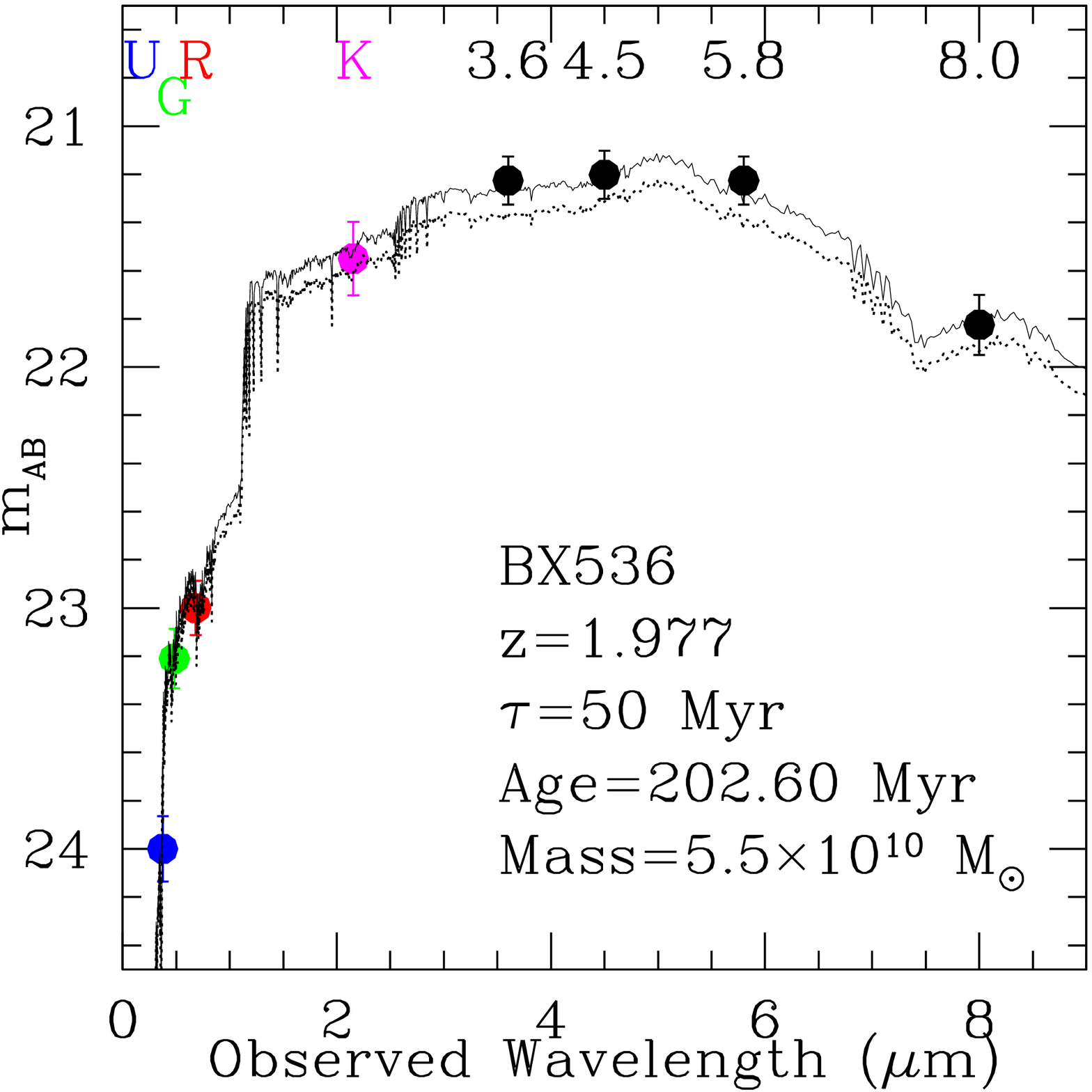,width=6.5cm}
\hspace{0.5cm}
\epsfig{figure=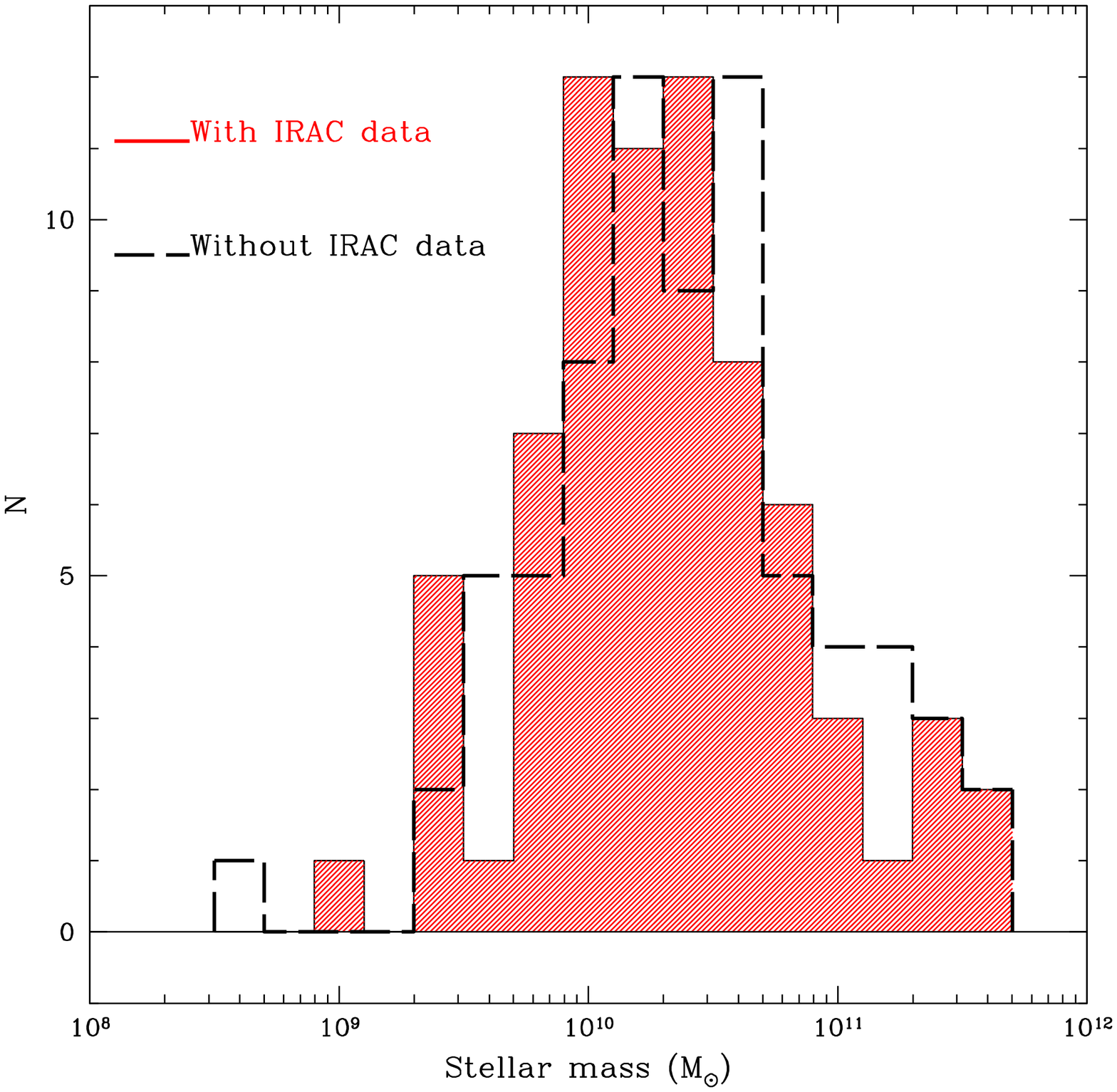,width=6.5cm}
}  
\end{center}
\caption{\emph{Left:} Model spectral energy distributions
(continuous lines) which best fit the rest-frame UV to near-IR 
photometry of Q1700-BX536, a typical star-forming galaxy at $z \simeq 2$. 
The bottom curve shows the result of fitting to the 
$UG{\cal R}K$ magnitudes only (that is, it excludes
the \emph{Spitzer}  IRAC data points). 
The values of the parameters which characterise the best-fitting
model---e-folding time scale of declining star formation, age, and 
assembled stellar mass---are given in the bottom right-hand corner.~
\emph{Right}: Histograms of assembled stellar masses for the 72
BX galaxies with spectroscopic redshifts in the field of
the $z_{\rm em} = 2.72$ QSO HS1700+643.
The median stellar mass of $z \simeq 2$ galaxies brighter
than ${\cal R} \simeq  25.5$ is $2 \times 10^{10} M_{\odot}$.
Both figures have been reproduced from Shapley et al. (2005).}
\end{figure}

The near-solar metallicities measured in star-forming
galaxies at $z = 2  - 3$---whether selected at optical, 
IR, or sub-mm wavelengths---are just one strand of a multi-faceted 
picture which has emerged in the last few years linking these
objects to the bulge component of today's galaxy population,
as originally proposed by Steidel et al. (1996). Their emission
properties at X-ray (Reddy \& Steidel 2004) and mid-IR 
(Reddy et al. 2006) wavelengths indicate star formation rates
${\rm SFR} \ga 10 M_{\odot}$~yr$^{-1}$ 
(for galaxies brighter than ${\cal R} = 25.5$) and
an average bolometric luminosity 
$\langle L_{\rm bol} \rangle \simeq 3 \times 10^{11} L_{\odot}$.
Their spectral energy distributions, which have now been
determined from the rest-frame
UV to the near-IR thanks to the highly
successful \emph{Spitzer} Space Telescope, indicate that these
high star formation rates typically persist for several hundred
million years, leading to the rapid assembly of stellar masses
$M_{\ast}  \ga 10^{10} M_{\odot}$ (see Figure 8). 
In models of galaxy formation which couple cosmological
simulations of dark matter haloes with the
physics and dynamics of the baryons (e.g. Blaizot et al. 2004;
Mori \& Umemura 2006) objects exhibiting these properties
at $z = 2 - 3$ naturally evolve to become today's massive 
and metal-rich elliptical galaxies.

%
%
\begin{figure}[h] 
\begin{center}
\vspace*{-0.2cm}
{\hspace*{-0.5cm} \epsfig{figure=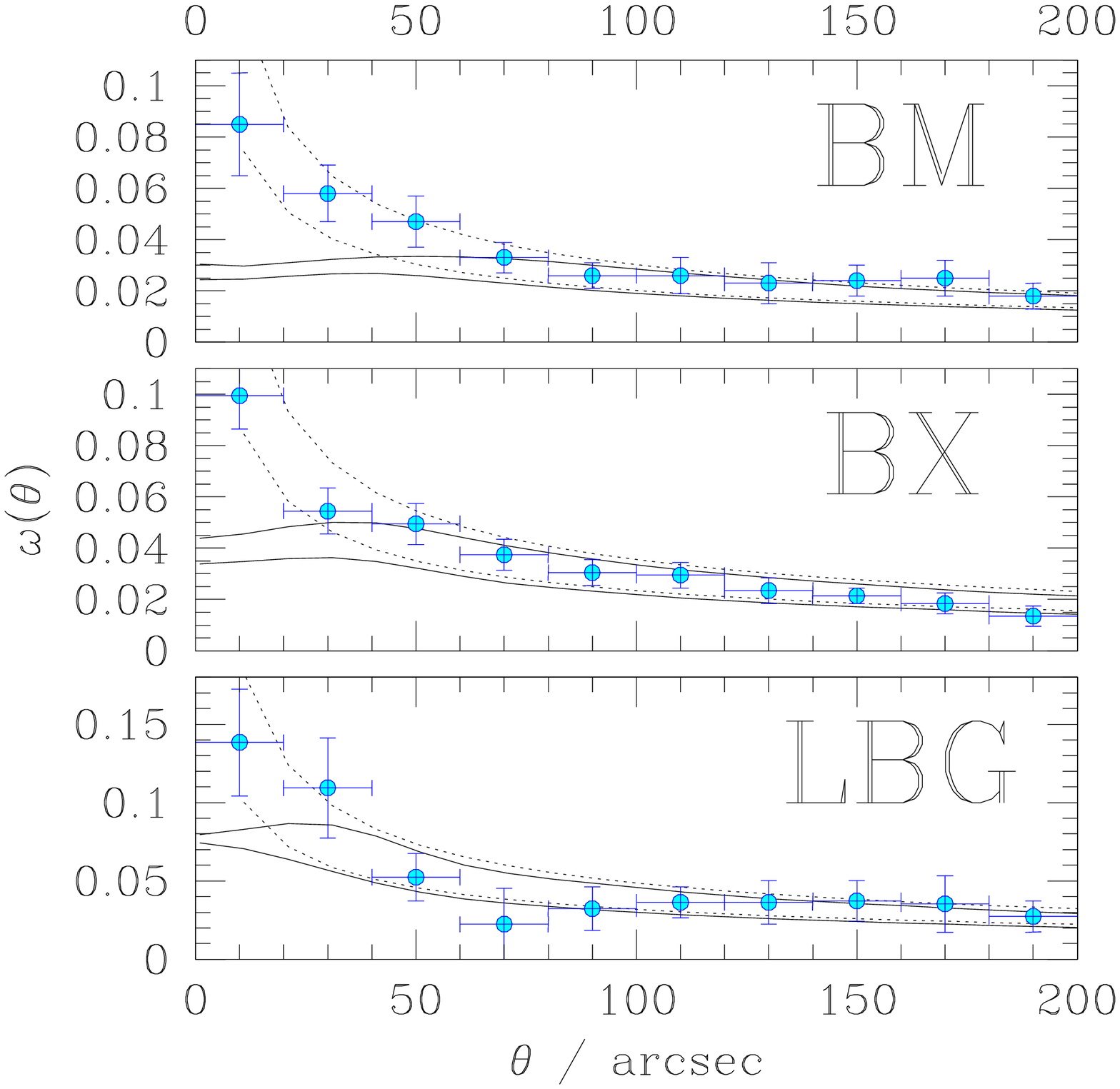,width=7cm}
\hspace{0.75cm}
\epsfig{figure=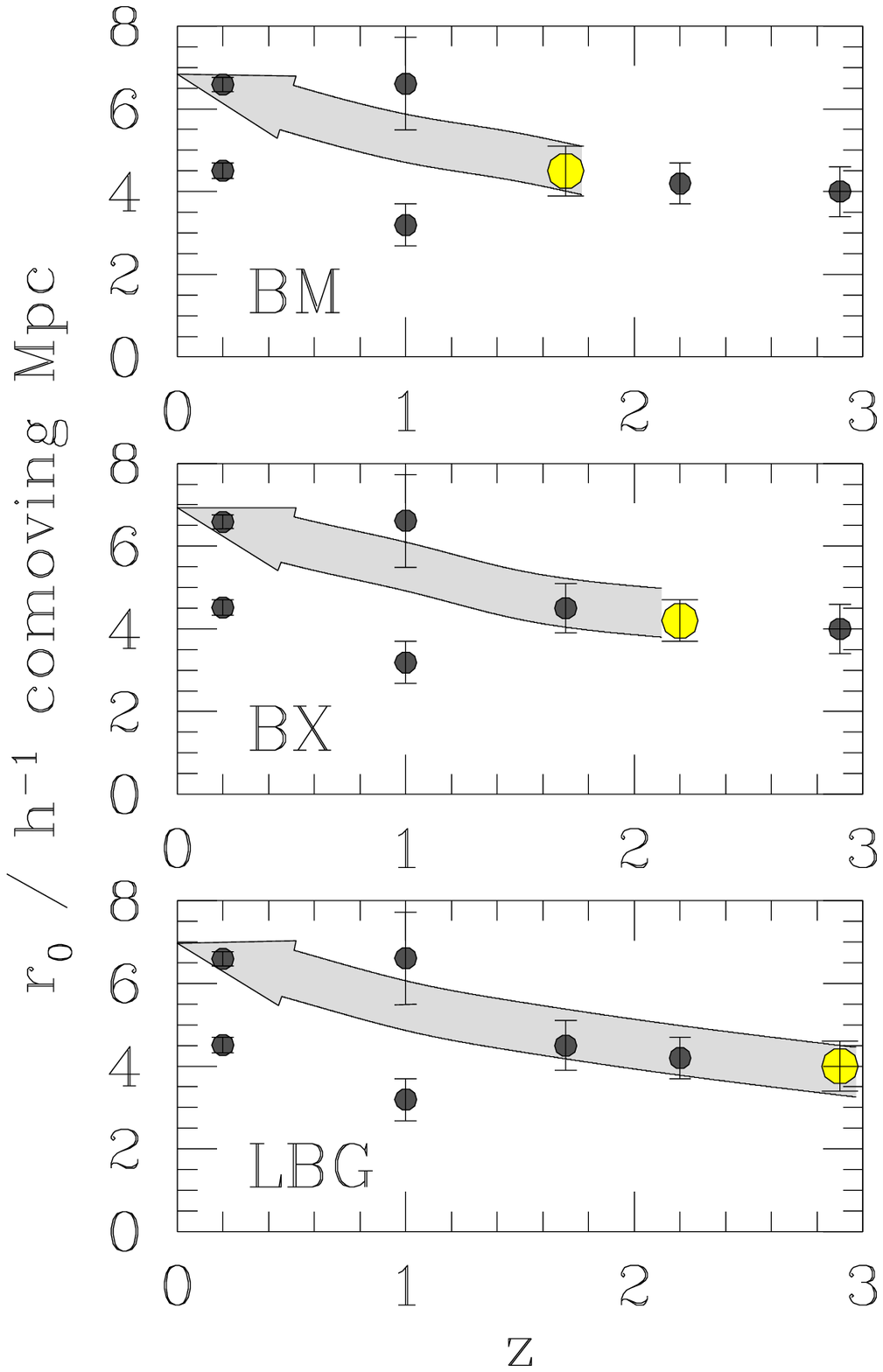,width=4.35cm}
}  
\end{center}
\vspace{-0.25cm}
\caption{\emph{Left}: Angular clustering on the sky of `BM'  
($\langle z \rangle = 1.7$), `BX' ($\langle z \rangle = 2.2$),
and Lyman break ($\langle z \rangle = 3.0$) galaxies (data
points) and of dark matter halos from cosmological
simulations (continuous lines). The simulations match the
observations for halo masses in the range 
$M_{\rm DM} = 10^{11.2} - 10^{12.3} M_{\odot}$.~
\emph{Right}: The arrows show the evolution seen in the simulations
for halos within the mass ranges which match the observed
galaxy clustering. The BM, BX, and LBG samples
seem to trace the same objects---or, more precisely, the
same halo mass range---through different evolutionary stages.
When evolved to intermediate and low redshifts, their
clustering matches that of elliptical galaxies in the DEEP
and 2dF surveys (the two upper data points at $z = 1$ and 0.2),
rather than that of galaxies which are still forming stars at
these later epochs (lower data points).
Both figures have been reproduced from
Adelberger et al. (2005b).}
\end{figure} 

The closest, and best studied, spiral bulge is that of the Milky Way
where the properties of individual stars can be determined.
Most models which have been proposed to explain the distributions
of stellar ages, metallicities, and element ratios (e.g. Ferreras,
Wyse, \& Silk 2003) share the same basic ingredients: 
(a) an early epoch of formation, at $z \sim 5$; 
(b) short infall timescales, $\tau < 0.5$\,Gyr, leading to
a rapid enrichment to near-solar metallicity and enhanced
[$\alpha$/Fe] ratios; and (c) significant outflow of gas and metals
accompanying the collapse and star formation process, resulting
in a final stellar mass $M_{\ast} \sim 10^{10} M_{\odot}$.
These characteristics are very much in line with those determined
for a typical UV-selected (`BX') galaxy at $z \sim 2$, although galaxies which
\emph{end up} with an assembled stellar mass of only
$M_{\ast} \sim 10^{10} M_{\odot}$ are probably near, or below, 
the faint end of the observed luminosity distribution.

%
%
\begin{figure}[] 
\begin{center}
\vspace*{0.25cm}
{\hspace*{-0.25cm} \epsfig{figure=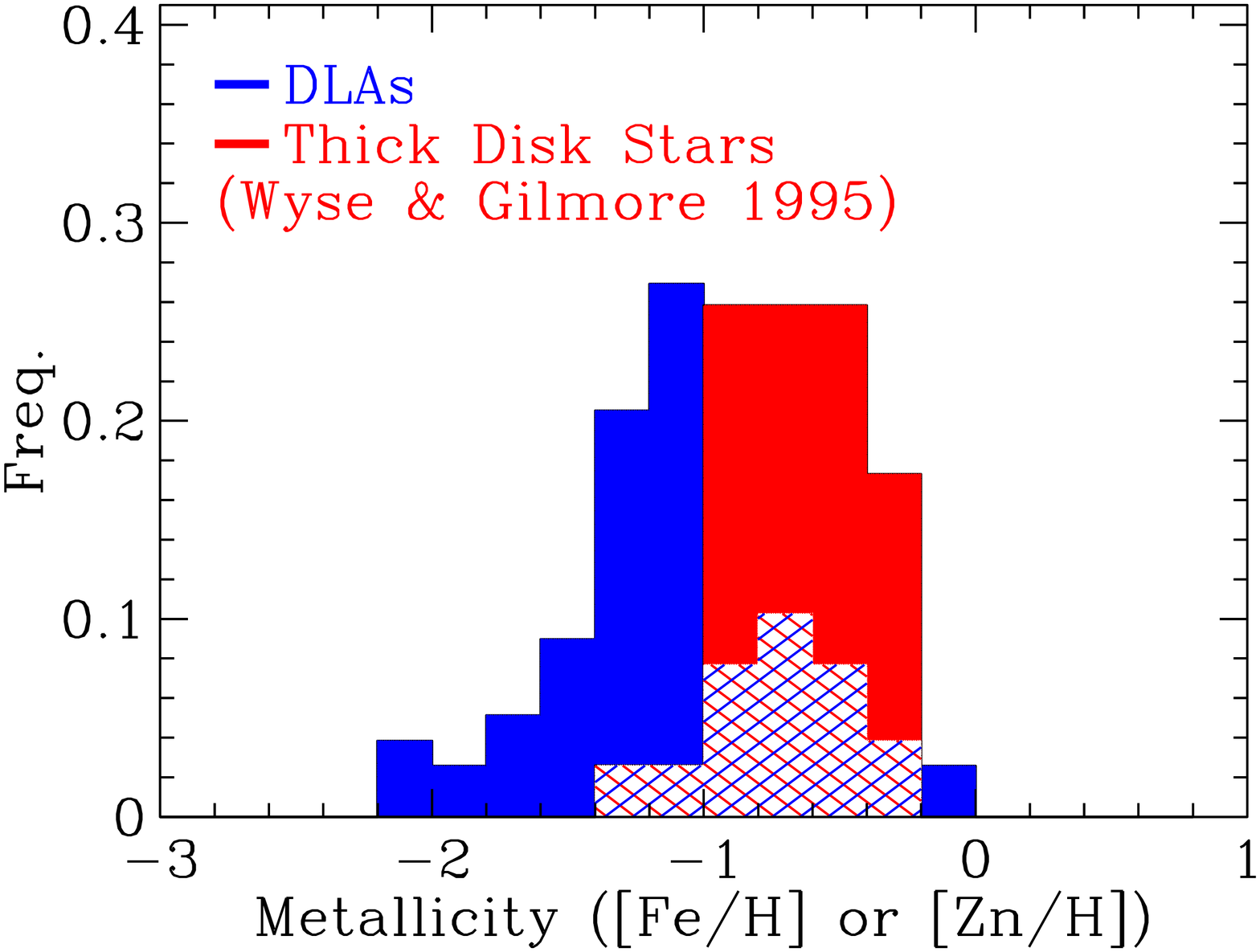,width=5.1cm,angle=270}
\hspace{0.65cm}
\epsfig{figure=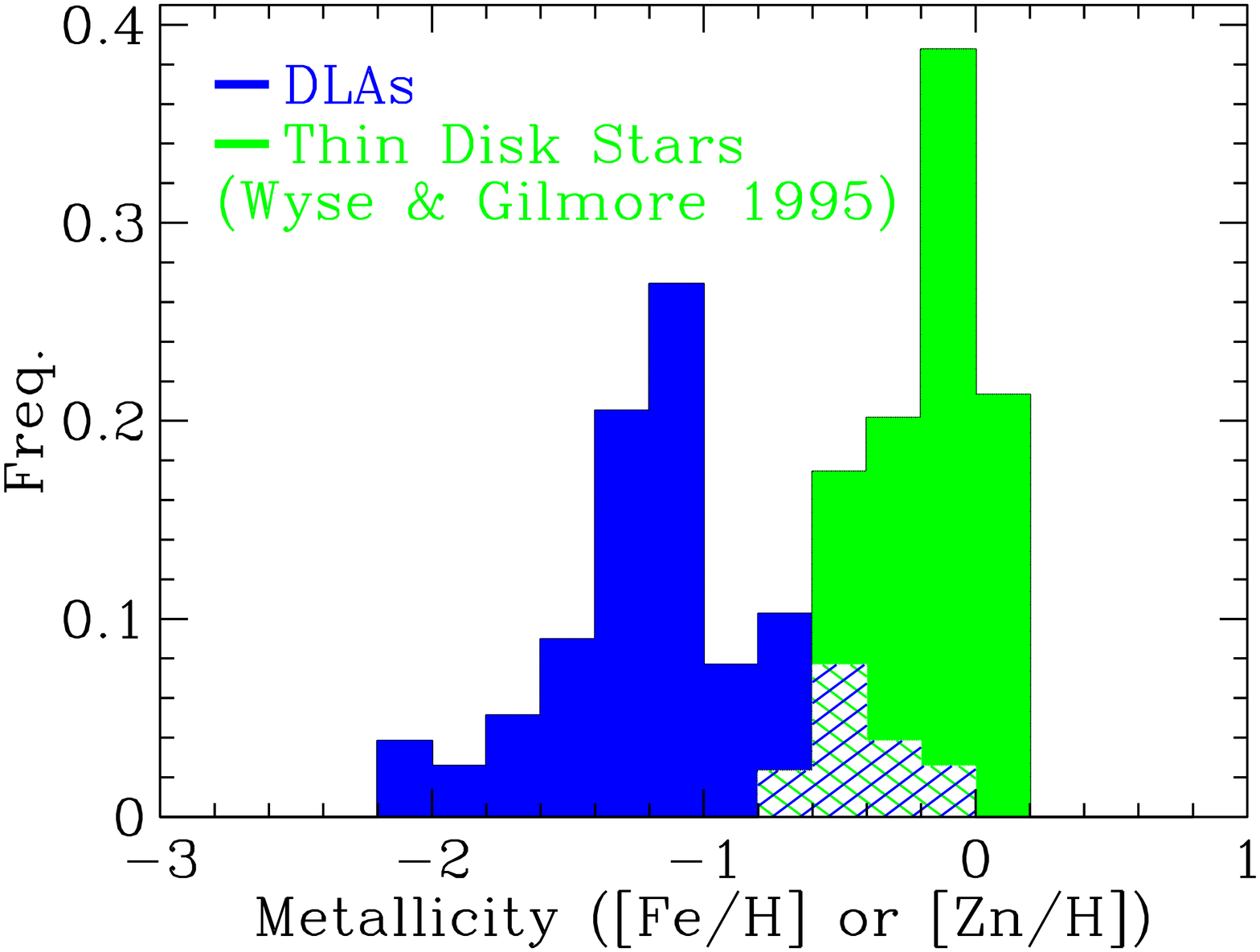,width=5.1cm,angle=270}
}  
\end{center}
\vspace{-0.15cm}
\caption{Metallicity distributions of DLAs and Milky Way thick (left panel)
and thin (right panel) disk stars
from the work by Akerman et al. (2005).
The metallicity is measured from Zn in DLAs and Fe in Galactic stars.
Approximately one half of the DLA measurements of [Zn/H] are upper limits;
they have been included in the histograms as if they were detections.
}
\end{figure} 

Perhaps the most compelling evidence linking the UV- and IR-bright
galaxies at $z = 2 - 3$ with today's elliptical galaxies is provided
by their clustering properties (see Figure 9). As shown by Adelberger
et al. (2005b), the high redshift populations are highly clustered,
with correlation lengths $r_0 = 4.0 - 4.5 h^{-1}$\,Mpc. 
In cosmological simulations, dark matter halos with these values 
of $r_0$ have masses $M_{\rm DM} = 10^{11.2} - 10^{12.3} M_{\odot}$;
evolved forward in time their clustering and number density 
are a much better match to those of elliptical galaxies than spirals.
There are also indications that in the largest galaxy overdensities
at $z \simeq 2$ star formation started earlier than in the field,
as expected within the general framework of hierarchical clustering
theories of the evolution of structure. As a consequence, galaxies
within these overdensities tend to have higher stellar masses and
metallicities than the mean values for the population as a whole
(Steidel et al. 2005).\\

\subsection{DLAs: Early Stages in the Formation of Spiral Galaxies?}

While an evolutionary 
link between star-forming galaxies at $z = 2 - 3$
and the spheroidal component of the galaxy population today
is strongly suggested by the available evidence,
an analogous connection for DLAs is still a matter for debate.
The working hypothesis that in DLAs we see the progenitors
of today's disk galaxies like the Milky Way (Wolfe et al. 1986),
while consistent with the kinematics of the associated
metal absorption lines (Wolfe et al. 2005),
runs into some difficulties when confronted with the metallicity
data (see Figure 10). 
Current models (e.g. Naab \& Ostriker 2006)
envisage the Milky Way disk to have been 
significantly smaller at $z = 2 - 3$ (that is, disk growth is from `inside out'),
and this would indeed explain why direct detection of 
the stellar light from DLA host galaxies at $z > 1$
has proved so elusive 
(Wolfe et al. 2005 and references therein).
However, even in the early stages of its evolution, the disk
was already more metal rich than most DLAs, with $Z \simeq 1/3Z_{\odot}$; 
the flat age-metallicity relation of disk stars implies that the epoch when
the metallicity of the Milky Way was below this value was brief
(e.g. Akerman et al. 2004).
Consequently, there is little overlap between the metallicity
distributions of the DLAs on the one hand, and of stars 
in the thin and thick disks of the Galaxy on the other
(Figure 10).

The low metallicities of most DLAs may be reconciled with the
proposed origin in spiral galaxies if, during their assembly, disk galaxies
were surrounded by extended envelopes (a halo, or a thick disk) of 
neutral gas which was metal-poor, or had a steep metallicity gradient.
Such a scenario has been proposed by Wolfe et al. (2003) and may
be the explanation for the very extended stellar disk structures 
recently found in galaxies such as M31 (Ibata et al. 2005). 
Simulations of the growth of disk galaxies through 
mergers of gas-rich subunits (e.g. Navarro 2004; Governato et al. 2005)
can reproduce many of the characteristics
of the Local Group of galaxies---it will be of considerable interest
to examine in more detail the properties of the gaseous components
of such mergers during this early build-up epoch with an eye to their 
possible interpretation as DLAs.

One of the most significant advances towards answering the long-standing 
question of which galaxies give rise to damped Ly$\alpha$ systems is the 
recent work by Zwaan et al. (2005) who tackled the problem 
`from the other end', as it were, by considering the H~{\sc i} properties 
of local galaxies. Based on 21\,cm column density
maps of nearly 400 galaxies, obtained as part of the Westerbork 
H~{\sc i} survey of spiral and irregular galaxies (WHISP)
and covering all Hubble types and a wide range
in luminosity, the analysis by Zwaan et al. has shown that
the distribution of luminosities of the galaxies producing DLAs
is nearly flat from $M_B = -21$ to $M_B = -15$. 
Thus, at $z = 0$,  galaxies spanning over two orders of magnitude
in luminosity evidently make roughly comparable contributions 
to the overall cross-section on the sky for DLA absorption. 
If this is also the case at high $z$,
the heterogeneous nature of DLA absorbers provides a plausible explanation
for the wide dispersion in their physical properties, including metallicity.\\

\section{Conclusions} 

\subsection{A Snapshot of Metallicity at $z = 2.5$}

The recent results summarised in this review have fleshed out further
the overall picture of element abundances at high redshift which has been
gradually coming into focus over the last few years. When we plot
the metallicities measured in different environments against the typical
physical scale of the structures to which they refer, as in Figure 11, 
we find a clear trend of decreasing metal abundance with increasing
scale. Another way to interpret the trend is to view the 
$x$-axis of Figure 11 as a scale of decreasing overdensity relative
to the cosmic mean. Thus, at $z = 2.5$, which corresponds to 
2.5\,Gyr after the Big Bang in today's consensus cosmology,
the gas in the inner 10--100\,pc of the highest overdensities, 
where supermassive black holes reside, has already been enriched
to supersolar proportions (e.g. Hamann et al. 2002). 
Galaxies which are undergoing active
star formation, at rates comparable to those of 
today's Luminous Infrared Galaxies, exhibit near-solar
metallicities on kpc scales. DLAs probably sample more diffuse gas,
possibly on the outskirts of galaxies in the process of formation and,
in any case, appear to be quite a heterogeneous population exhibiting
more than two orders of magnitude of dispersion in their degree of
metal enrichment. Finally,  the Lyman $\alpha$ forest which traces
large scale structures of moderate overdensity relative to
the cosmic mean (and which could not be included in this review due
to lack of space) contains only trace amounts of metals (whose
origin is still the matter of some debate---see Adelberger 2005a
and Songaila 2006).

%
%
\begin{figure}[h] 
\vspace*{-2.9cm}
\begin{center}
{\hspace*{0.3cm}\epsfig{figure=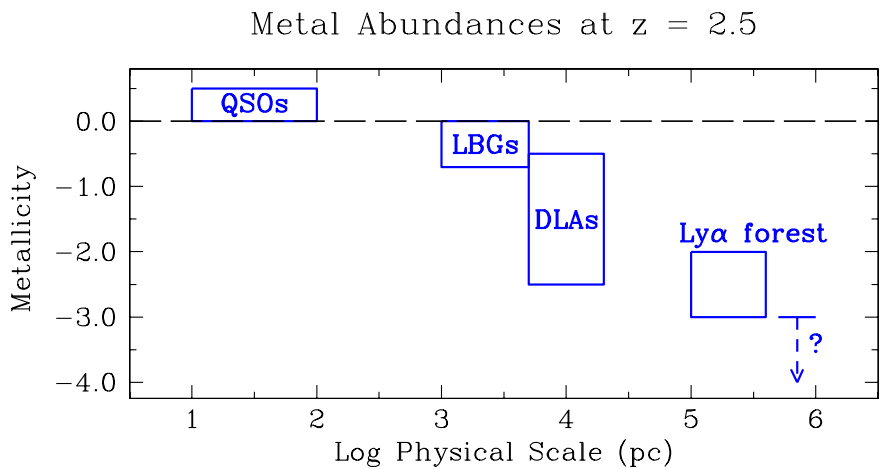,width=12.5cm,angle=270}}  
\end{center}
\vspace{-3.55cm}
\caption{Snapshot of the metallicity of different components of the
high redshift universe. The logarithm of the metallicity is plotted
relative to solar (indicated by the long-dash line at 0.0) against the 
typical linear scale of the structures to which it refers.
The term ``Lyman break galaxies'' (LBGs) is used as a shorthand here
to refer to a more general class of actively star-forming galaxies.
}
\vspace{0.25cm}
\end{figure} 

\noindent  Evidently, it is the depth of the potential well 
within which the
baryons find themselves that drives the pace at which 
gas is processed through stellar nucleosynthesis. 
Thus, at any one epoch, there can be a range of three-to-four 
orders of magnitude in the degree of metal enrichment 
which one would measure, depending on environment.\\

\subsection{Metallicity Evolution}

Conversely, metallicity depends weakly 
on cosmic time---the age-metallicity relation of the universe as a whole
appears to be shallow, just as is the case for the stellar 
disk of the Milky Way. 
This is the conclusion reached by the first attempts to assess
the degree of evolution with look-back time of relationships 
which are well established in the present-day universe, such 
as the luminosity-metallicity and mass-metallicity relations.
The obstacles faced by such endeavours are considerable.  
One is the obvious requirement that we should
employ measures of metallicity which can be applied 
self-consistently over a wide range of redshifts.
This is not straightforward
to implement, as explained in the preceding sections, 
but in principle should be a tractable problem given
sufficient telescope time and attention to detail.
Far less clear is how to isolate some meaningful
property which we can hold constant with redshift 
and against which metallicity is to be measured---in
an evolving universe how can we be sure that
we are comparing the same kinds of objects at different
epochs?

One of the first attempts in this direction is that 
by Kewley \& Kobulnicky (2005) who deduced a slope 
of $-0.15$\, dex per unit redshift in the oxygen abundance
of luminous galaxies, brighter than $M_B = -20.5$ or approximately
$L^{\ast}$ (see also Maier, Lilly \& Carollo 2005). 
A similarly shallow slope is exhibited by the 
redshift evolution of the metallicity of DLAs (Kulkarni et al. 2005).
In other words, from $z = 3$ to 0
the average (O/H) increased by a mere factor of three: this
is modest evolution indeed, compared with the
orders-of-magnitude dispersion at a given epoch seen
in Figure 11. 
The difficulty lies in the interpretation of $M_B$, given that we
are most likely dealing with objects of very different
mass-to-light ratios at different redshifts, as shown by
Shapley et al. (2005).

%
%
\begin{figure}[]
\begin{center}
{\hspace*{-0.4cm} \epsfig{figure=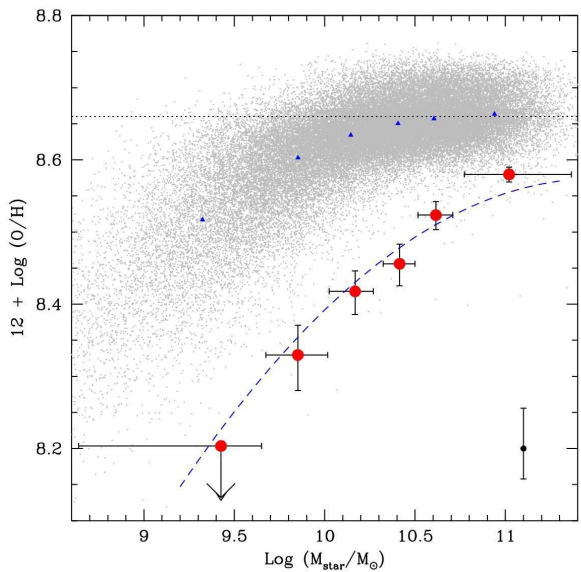,width=8cm}}  
\end{center}
\vspace{-0.3cm}
\caption{Stellar mass-metallicity relation for star-forming galaxies
at $z \sim 2$ (large red circles with error bars) and in the nearby
($z \sim 0.1$) universe from the SDSS survey (small grey points). 
The blue triangles show the mean metallicity of the SDSS galaxies 
in the same mass bins as the $z = 2$ galaxies. The vertical error bar
in the bottom right-hand corner indicates the uncertainty in the calibration
of the $N2$ index used to deduce the oxygen abundance of both
sets of galaxies. Since the $N2$ index saturates near solar metallicity
(indicated by the horizontal dotted line), the offset between the 
present-day and high-redshift relations is best derived by consideration
of the lower metallicity bins. At a given assembled stellar mass, 
galaxies at  $z = 2$ appear to be about a factor of two lower 
in metallicity than today.
(Figure reproduced from Erb et al. 2006a).}
\end{figure} 

In this respect mass is probably a more 
physically meaningful quantity than light.
While we cannot determine directly the masses of the dark matter
halos in which \emph{individual} galaxies reside (the
resolution and sensitivity required to, for example, trace 21\,cm rotation curves
at high redshift are still years away), we do have estimates of the 
assembled stellar mass---the integral of the past star formation rate
over the age of the galaxies---as discussed in section 4.1.
When compared with the present-day mass-metallicity relation
(see Figure~12), we find again only a modest degree of evolution:
galaxies of a given stellar mass are on average about a factor of two
lower in metallicity at $z \simeq  2$ than today, consistent with the
shallow metallicity-redshift gradient deduced by Kewley \& Kobulnicky
(2005) and by Kulkarni et al. (2005).
The more significant shift in Figure~12 is along the $x$-axis, as 
pointed out by Savaglio et al. (2005): galaxies which had
attained a given metallicity were one order of magnitude 
more massive at $z \simeq 2$.\footnote{
This statement is correct only if the relationship between
metallicity and gas fraction does not change significantly
among galaxies of different masses and at different redshifts.
Clearly this is an oversimplification, only applicable to
a `closed box' model of galactic chemical evolution with a universal
IMF. Nevertheless, if variations in the effective yield---which links
metallicity and gas fraction---are not extreme, then the above
conclusion is valid to a first approximation.}
This `evolution' is in the sense of 
the conclusions we drew from Figure 11, reinforcing the idea
that the cycling of gas through stars proceeds at a pace determined
primarily by the overdensity of the objects which have collapsed to
form bound structures.\\

\subsection{A Revised Census of Metals at $z = 2.5$}

The recent advances in determining the 
physical properties of galaxies at high redshift
summarised in this review 
warrant revisiting the `Missing Metals' problem
discussed by Pagel (2002) and Pettini (2004), and more recently
reconsidered by Bouch\'{e}, Lehnert, \& P\'{e}roux (2005).
Briefly, the (unobscured) ultraviolet luminosity of galaxies
is also a measure of the metal production rate, since the same
massive stars produce the UV photons and the metals which are
promptly released into the ISM.
One can thus integrate the comoving star formation rate deduced
from galaxy surveys and, with appropriate conversion factors,
obtain an estimate of the comoving density of metals which have
accumulated in the universe by a given redshift. 
A simple consistency check is to compare this quantity with
the metallicity of different components of the high redshift universe
and thereby obtain an indication of how (in)complete our
census of metals at high $z$ is. 
Initial indications (Pagel 2002; Pettini 2004) pointed to a 
significant deficit in the metals accounted for by LBGs, DLAs
and the Lyman $\alpha$ forest, which we are now ready to 
reexamine with improved data.

I begin by recalculating the total metal production up to $z = 2.5$.
Recent studies of the X-ray (Reddy \& Steidel 2004) and mid-IR
(Reddy et al. 2006) luminosities of (rest-frame) 
UV-selected galaxies at $z = 2 - 3$,
making use of the very deep imaging at many wavelengths 
of the GOODS-N field, have placed on more secure footing 
estimates of the extinction correction applicable to UV-deduced
star formation rates. The typical correction factor of 4.5-5 obtained
by Reddy \& Steidel (2004) is in excellent agreement with 
the value of 4.7  adopted by Steidel et al. (1999)
from consideration of the observed UV spectral slope 
of Lyman break galaxies. Furthermore, the GOODS-N field
also allowed different photometric selections of high $z$ galaxies
to be compared quantitatively. Reddy et al. (2005) considered
the contribution of optical, near-IR, and sub-mm selected galaxies
to the star formation rate density at $z \sim 2$, taking into
account sample overlap, and concluded that optically (rest-frame UV)
selected galaxies brighter than ${\cal R} = 25.5$ (the limit of the 
surveys by Steidel and collaborators) account for $\sim 70$\% 
of the total. 
Star formation and metal ejection rates are related by
$\dot{\rho}_{\ast} = 64 \, \dot{\rho}_{Z}$  according to
the recent estimate by Conti et al. (2003); this conversion factor
is a reduction by a factor of 1.5 in the metal production
rate compared to the original estimate by Madau et al. (1996).
With these updates, and converting the estimates of the
star formation rate density by Steidel et al. (1999) to today's
consensus cosmology 
($\Omega_{\rm M} = 0.3$, $\Omega_{\Lambda} = 0.7$, 
$H_0 = 70$\,km~s$^{-1}$~Mpc$^{-1}$),
we obtain:
\begin{equation}
	\int_{11\,Gyr}^{13\,Gyr}\dot{\rho}_{Z}~dt \simeq
	3.4 \times 10^6\,M_{\odot}~{\rm Mpc}^{-3} ~. 
\end{equation}
This value is obtained under the assumption that star formation in the universe
began at $z = 10$ (0.46\,Gyr after the Big Bang) at levels similar to those
measured at $z = 4 - 5$. An increasing star formation rate density 
from $z = 10$ to 4, as proposed by 
Bouwens et al. (2005---see also these proceedings) 
and Bunker et al. (2005),
would reduce the above estimate by about one quarter.

A comoving density of metals of 
$3.4 \times 10^6\,M_{\odot}~{\rm Mpc}^{-3}$
corresponds to
\begin{equation}
   \Omega_Z \simeq 0.045 \times {\rm (}\Omega_{\rm baryons} \times 0.0126 {\rm )}
   \label{}
\end{equation}
where  
$\Omega_{\rm baryons} = 0.044$  (Pettini 2005) of 
$\rho_{\rm crit} = 1.35 \times 10^{11} \, M_{\odot}~{\rm Mpc}^{-3}$
(both for $h_{70} = 1$)
and 0.0126 is the mass fraction of elements heavier than helium
for solar metallicity (Asplund et al. 2004). The value of $\Omega_Z$ 
obtained here is not very different from that calculated by Pettini (2004)
because the various updates to the measurements on which it is based
largely compensate each other. It implies that the star formation activity
we believe to have taken place between the Big Bang and $z = 2.5$
(2.5\,Gyr later) was sufficient to enrich the universe as whole to a 
metallicity of 4--5\% of solar. 

%
%
\begin{figure}[h]
\vspace*{-7.25cm}
\hspace*{-0.7cm}
\psfig{figure=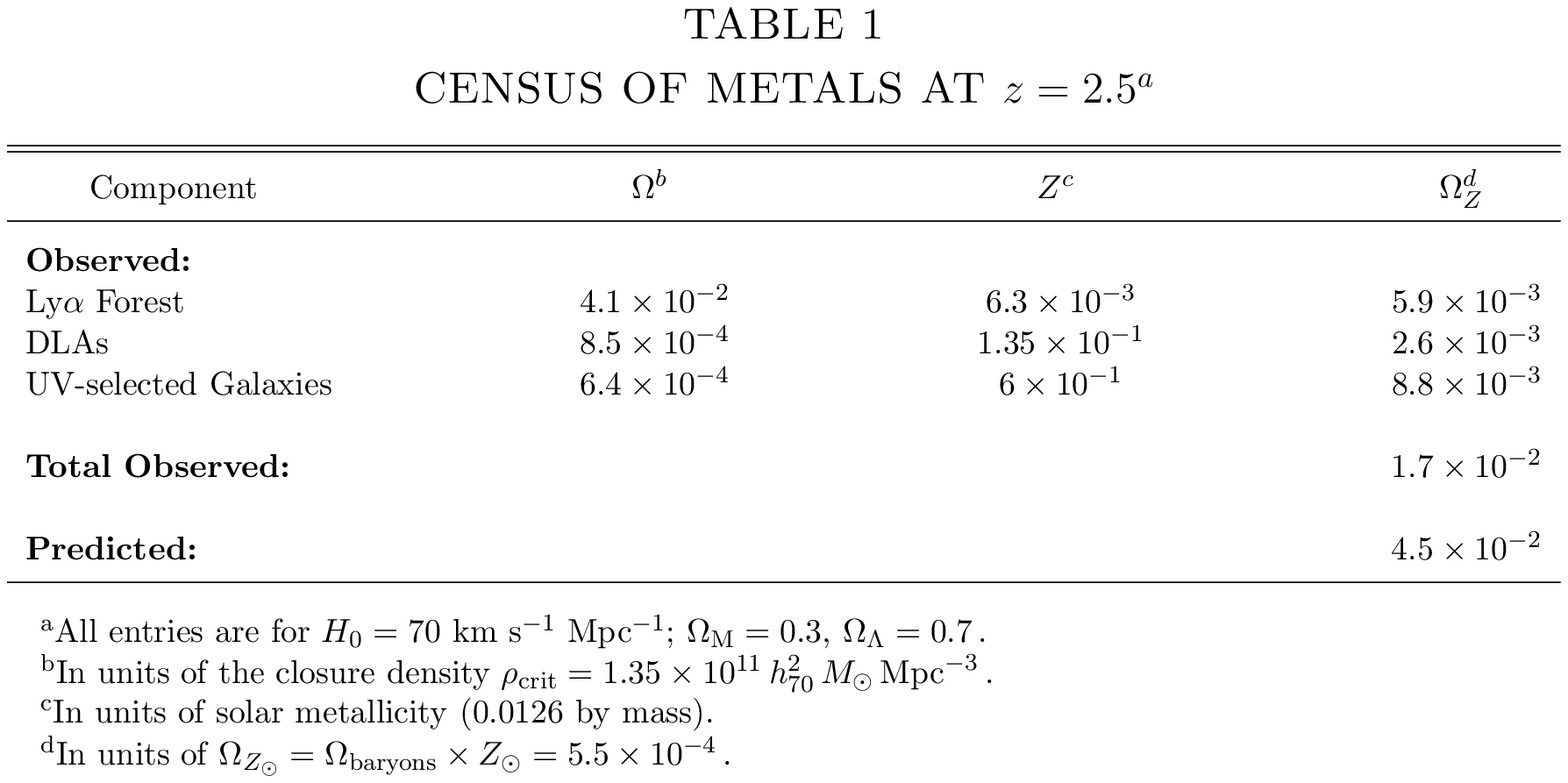,width=17.5cm,angle=0}
\vspace{-8.75cm}
\end{figure}

\noindent Table 1 assesses the contributions of the 
Lyman~$\alpha$ forest, DLAs and UV-selected galaxies to the total metal budget.
For the Lyman~$\alpha$ forest I quote the results of the
careful study by Simcoe, Sargent, \& Rauch (2004),
based on associated C~{\sc iv} and O~{\sc vi}
absorption in QSO spectra of very high S/N.
For DLAs,  $\Omega_{\rm DLA} = 8.5 \times 10^{-4}$ is
indicated by the improved statistics
of the SDSS (Prochaska, Herbert-Fort, \& Wolfe 2005), and 
$Z_{\rm DLA} = 2/15 Z_{\odot}$ is the value implied by 
the column density-weighted metallicity 
$[ \langle {\rm Zn/H} \rangle] = -0.87 \pm 0.13$
of Akerman et al. 2005 (no correction has been applied for a possible
underabundance of the Fe-peak elements relative to elements
which are released promptly into the ISM because such an effect
is generally not seen in DLAs---see Figure 7).
When added together, QSO absorbers (the Lyman~$\alpha$ forest and
DLAs) make up less than 20\% of the required 
$\Omega_Z \simeq 0.045$ (last column of Table 1);
interestingly, these recent estimates indicate
that there are approximately twice as many metals in 
the forest as in DLAs.
Lyman limit systems (sometime referred as 
sub-DLAs) have not been included in our accounting, 
because their comoving mass density
and metallicity have not yet been established with certainty.
If the gas in sub-DLAs is mostly ionised, 
their contribution to the metal budget
could be significant (P{\'e}roux et al. 2005).
We could also be missing metals in the Lyman~$\alpha$
forest, if they are in a hot phase which does not give rise to
detectable C~{\sc iv} and O~{\sc vi} absorption lines. 

Let us now turn to UV-selected galaxies. Adelberger \& Steidel (2000)
showed that the luminosity function of LBGs at $z = 3$ is well fitted
by a Schechter function with faint end slope $\alpha = -1.6$,
characteristic magnitude $m_{\ast}({\cal R}) = 24.54$, 
and normalisation 
$\Phi_{\ast} = 1.5 \times 10^{-3} h_{70}^{3}$\,Mpc$^{-3}$.  
These parameters do not appear
to change from $z \simeq  4$ (Steidel et al. 1999) to $z \simeq 2$
(Reddy et al. in preparation), 
apart from the scaling of $m_{\ast}({\cal R})$
in accordance to the luminosity distance and $k$-corrections
applicable to the different redshifts.
Thus, the comoving density
of UV-selected galaxies brighter than ${\cal R} = 25.5$
($\approx 1/4 m_{\ast}$) at
$z = 2.2$ (the median redshift of the BX sample to which the
metallicity measurements discussed in \S3.1 refer) is
$2.4 \times 10^{-3}$\,Mpc$^{-3}$ .

The work of Erb et al. (2006a) has provided estimates
of the baryonic mass (stars+gas) and oxygen abundance
for a sample of nearly one hundred such galaxies:
masses range between $\log (M/M_{\odot}) = 11.1$ and
10.3, and metallicities between approximately solar to less than
1/3 solar.
A proper accounting would integrate the mass-weighted
metallicity of these galaxies; 
however, such an approach would require a knowledge
of the mass function of BX galaxies, which has yet to be determined.
Given the relatively small range of baryonic masses and metallicities
indicated by the analysis of Erb et al. (2006a),
we may be justified in adopting the median values of the 
sample as a first approximation. Thus, combining
the median baryonic mass of BX galaxies
$\langle  M_{\rm BX} \rangle = 3.6 \times 10^{10} \, M_{\odot}$
with their comoving volume density given above,
we obtain a total baryonic mass density
$\Omega_{\rm BX} =  6.4 \times 10^{-4}$.
With a typical metallicity 
$Z_{\rm BX} \simeq 0.6 Z_{\odot}$, the metal content
of UV-selected galaxies amounts to 
$\Omega_{Z} =  8.8 \times 10^{-3}$.

In reality, this is likely to be an underestimate because the
data of Erb et al. show a correlation between mass and metallicity
at the high mass end (see Figure 8 of Erb et al. 2006a),
so that the more massive galaxies contain proportionally more metals.
Furthermore, we have not included in our accounting
galaxies which are missed by the UV photometric
selection technique but included in near-IR and sub-mm surveys.
To date the masses and metallicities of these objects are less
well determined than those of UV-selected galaxies.
These different samples of high redshift galaxies
do overlap, as shown by Reddy
et al. (2005) who concluded that the galaxies missed by
the UV selection account for $\sim 30$\% of the star
formation rate density at $z \simeq 2$. However, such galaxies may
contribute a larger fraction of the metals,
if they include a high proportion of
evolved objects which have attained a high metallicity
(e.g. Dunne, Eales, \& Edmunds 2003).

If we add up the contributions of the Lyman~$\alpha$ forest,
DLAs and UV-selected galaxies (last column of Table~1),
we can account for $\sim 40\%$ of the metal
production from the Big Bang to $z = 2.5$. 
With the addition of metals in sub-DLAs, in 
the most massive UV-selected galaxies, and in galaxies
missed by the BX photometric criteria---all unaccounted
for in the estimates above---we may well
have solved the `Missing Metals' problem.
Perhaps this is an indication that our knowledge of the
high redshift galaxy population has indeed improved
in the last few years.

\acknowledgements{It is a pleasure to acknowledge my collaborators
in the various projects described in this review, particularly Kurt Adelberger, 
Chris Akerman, David Bowen, Sara Ellison, Dawn Erb, Naveen Reddy, 
Sam Rix, Alice Shapley and Chuck Steidel. I am indebted to Bernard Pagel for 
numerous inspiring conversations over the years, and I am grateful
to him, Miroslava Dessauges-Zavadsky, Varsha Kulkarni, Claus Leitherer, and 
Christy Tremonti for valuable comments on this paper.
Finally, I thank the conference organisers for financial support which helped
me take part in this stimulating meeting.
}

\vfill 

\begin{thebibliography}{}{ 

\bibitem{} Adelberger, K.~L. 2005a, in Probing Galaxies through
Quasar Absorption Lines, IAU Colloquium 199, 
P.~R. Williams, C. Shu, \& B. M\'{e}nard eds.
(Cambridge: Cambridge University Press), 341

\bibitem{} Adelberger, K.~L. \& Steidel, C.~C. 2000, ApJ, 544, 218

\bibitem{} Adelberger, K.~L., Steidel, C.~C., Pettini, M.,
Shapley, A.~E., Reddy, N.~A., \& Erb, D.~K. 2005b, ApJ, 619, 697

\bibitem{} Adelberger, K.~L., Steidel, C.~C., 
Shapley, A.~E., Hunt, M.~P., Erb, D.~K., Reddy, N.~A., \& 
Pettini, M.\  2004, ApJ, 607, 226 

\bibitem{} Akerman, C.~J., Carigi, L., Pettini, M., Nissen, P.~E. \& Asplund, M.
2004, A\&A, 414, 931

\bibitem{} Akerman, C.~J., Ellison, S.~L., Pettini, M., \& Steidel, C.~C. 2005,
A\&A, 440, 499

\bibitem{} Asplund, M., Grevesse, N., Sauval, A.~J., Allende Prieto, C.,
\& Kiselman, D. 2004, A\&A, 417, 751

\bibitem{} Blaizot, J., Guiderdoni, B., Devriendt, J.~E.~G., Bouchet,
F.~R., Hatton, S.~J., \& Stoehr, F. 2004, MNRAS, 352, 571

\bibitem{} Bouch\'{e}, N., Lehnert, M.~D., \& P\'{e}roux, C. 2005,
MNRAS, 364, 319

\bibitem{} Bouwens, R.~J., 
Illingworth, G.~D., Thompson, R.~I., \& Franx, M.\ 2005, \apjl, 624, L5 

\bibitem{} Bowen, D.~V., Jenkins, E.~B., Pettini, M., \& Tripp, T.~M. 2005,
ApJ, 635, 880

\bibitem{} Bunker, A., Stanway, E., Ellis, R., McMahon, R., Eyles, L., \& Lacy, M.
2005, in First Light and Reionization, E. Barton \& A. Cooray 
eds., New Astronomy Reviews, in press (astro-ph/0508271)

\bibitem{} Cannon, J.~M., Skillman, E.~D., Sembach, K.~R., \& Bomans, D.~J. 2005,
ApJ, 618, 247

\bibitem{} Conti et al. 2003, AJ, 126, 2330

\bibitem{} de Mello, D.~F., 
Daddi, E., Renzini, A., Cimatti, A., di Serego Alighieri, S., Pozzetti, L., 
\& Zamorani, G.\ 2004, ApJ, 608, L29 

\bibitem{} Dunne, L., Eales, S.~A., 
\& Edmunds, M.~G.\ 2003, \mnras, 341, 589 

\bibitem{} Ellison, S.~L., Yan, L., Hook, I.~M., Pettini, M., 
Wall, J.~V., \& Shaver, P.  2001, A\&A, 379, 292

\bibitem{} Erb, D.~K., Shapley, A.~E., Pettini, M., Steidel, C.~C.,
Reddy, N.~A., \& Adelberger, K.~L. 2006a, ApJ, in press (astro-ph/0602473) 

\bibitem{} Erb, D.~K., Steidel, C.~C., Shapley, A.~E., Pettini, M., 
Reddy, N.~A., \& Adelberger, K.~L. 2006b, ApJ, submitted

\bibitem{} Ferreras, I., Wyse, R.~F.~G. \& Silk, J. 2003, MNRAS, 345, 1381

\bibitem{} Garnett, D.~R., Bresolin, F., \& Kennicutt, R.~F. 2004, ApJ, 607, L21

\bibitem{} Governato, F., Stinson, G., Wadsley, J., \& Quinn, T. 2005, in
The Fabulous Destiny of Galaxies: Bridging Past and Present, 
in press (astro-ph/0509263)

\bibitem{} Hamann, F., Korista, 
K.~T., Ferland, G.~J., Warner, C., \& Baldwin, J.\ 2002, ApJ, 564, 592 

\bibitem{} Ibata, R., Chapman, S., Ferguson, A.M.N., Lewis, G.,
Irwin, M., \& Tanvir, N. 2005, ApJ, 634, 287

\bibitem{} Kewley, L., \& Kobulnicky, H.~A. 2005, in 
Starbursts: From 30 Doradus to Lyman Break Galaxies,
R. de Grijs \& R.~M. Gonz\'{a}lez Delgado eds.,
ASSL Vol.~329, (Dordrecht: Springer), 307 

\bibitem{} Kulkarni, V.~P. \& Fall, S.~M. 2002, ApJ, 580, 732

\bibitem{} Kulkarni, V.~P., Fall, S.~M., 
Lauroesch, J.~T., York, D.~G., Welty, D.~E., Khare, P., \& Truran, 
J.~W.\ 2005, ApJ, 618, 68 

\bibitem{} Leitherer, C., Le\~ao, J.~R.~S., Heckman, T.~M., Lennon, D.~J., 
Pettini, M., \& Robert, C.  2001, ApJ, 550, 724

\bibitem{} Leitherer, C., Schaerer, D., Goldader, J.~D., Delgado, R.~M.~G.,
Robert, C., Kune, D.~F., de Mello, D.~F., Devost, D., \& Heckman, T.~M. 1999,
ApJS, 123, 3

\bibitem{} Madau, P., Ferguson, H.~C., Dickinson, M.~E., Giavalisco, M.,
Steidel, C.~C., \& Fruchter, A. 1996, MNRAS, 283, 1388

\bibitem{} Maier, C., Lilly, S.~J., \& Carollo, C.~M. 2005, in 
The Fabulous Destiny of Galaxies: Bridging Past and Present, 
in press (astro-ph/0509114)

\bibitem{} Matteucci, F., \& Pipino, A. 2002, ApJ, 569, L69

\bibitem{} Matteucci, F., \& Recchi, S. 2001, ApJ, 558, 351

\bibitem{} Mehlert, D., et al.  2002, A\&A, 393, 809

\bibitem{} Mori, M., \& Umemura, M. 2006, Nature, in press (astro-ph/0512424)

\bibitem{} Naab, T., \& Ostriker, J.~P. 2006, MNRAS, 366, 899

\bibitem{} Nagamine, K., Springel, V., \& Hernquist, L. 2004,
MNRAS, 348, 435

\bibitem{} Navarro, J.~F.  2004, in Penetrating Bars through Masks
of Cosmic Dust, in press (astro-ph/0405497)

\bibitem{} Nissen, P.~E., Chen, Y.~Q., Asplund, M., \& Pettini, M.\ 2004, \aap, 415, 993 

\bibitem{} Pagel, B.~E.~J. 2002, in ASP Conf. Series 253, Chemical Enrichment
of Intracluster and Intergalactic Medium, R. Fusco-Femiano, \& F. Matteucci eds.
(San Francisco: ASP), 489

\bibitem{} Pagel, B.~E.~J., Edmunds, 
M.~G., Blackwell, D.~E., Chun, M.~S., \& Smith, G.\ 1979, MNRAS, 189, 95 

\bibitem{} Pagel, B.~E.~J., \& Tautvaisiene, G.\ 1995, \mnras, 276, 505 

\bibitem{} Pauldrach, A.~W.~A., Hoffmann, T.~L., \& Lennon, M.  2001, 
A\&A, 375, 161

\bibitem{} Pei, Y.~C., \& Fall, S.~M. 1995, ApJ, 454, 69

\bibitem{} P{\'e}roux, C., 
Dessauges-Zavadsky, M., D'Odorico, S., Sun Kim, T., \& McMahon, R.~G.\ 
2005, \mnras, 363, 479 

\bibitem{} Pettini, M. 2004, in Cosmochemistry: The Melting Pot of
Elements, C. Esteban, A. Herrero, R. Garcia-Lopez, \& F. Sanchez
eds. (Cambridge: Cambridge University Press),  257 (astro-ph/0303272)

\bibitem{} Pettini, M. 2005, in Astrophysics in the Far Ultraviolet,
G. Sonneborn, H.~W. Moos, \&  B.~G Andersson  eds. 
(San Francisco: ASP), in press (astro-ph/0601428)

\bibitem{} Pettini, M., Boksenberg, A., \& Hunstead, R.~W. 1990, ApJ, 348, 48

\bibitem{} Pettini, M., \& Pagel, B.~E.~J. 2004, MNRAS, 348, L59 

\bibitem{} Pettini, M., Rix, S.~A., Steidel, C.~C., Adelberger, K.~L.,
Hunt, M.~P., \& Shapley, A.~E. 2002, ApJ, 569, 742

\bibitem{} Prochaska, J.~X., Herbert-Fort, S. \& Wolfe, A.~M.  2005,
ApJ, 635, 123

\bibitem{} Reddy, N.~A., Erb, D.~K., Steidel, C.~C., Shapley, A.~E.,
Adelberger, K.~L., \& Pettini, M. 2005, ApJ, 633, 748i

\bibitem{} Reddy, N.~A., \& Steidel, C.~C. 2004, ApJ, 603, L13

\bibitem{} Reddy, N.~A., Steidel, C.~C., Fadda, D., Yan, L., 
Pettini, M., Shapley, A.~E., Erb, D.~K., \& Adelberger, K.~L.  2006,
ApJ, in press (astro-ph/0602596)

\bibitem{} Rix, S.~A., Pettini, M., Leitherer, C., Bresolin, F.,
Kudritzki, R., \& Steidel, C.~C. 2004, ApJ, 615, 98

\bibitem{} Savaglio, S., et al. 2004, ApJ, 602, 51

\bibitem{} Savaglio, S., et al. 2005, ApJ, 635, 260

\bibitem{} Schulte-Ladbeck, R.~E., Rao, S.~M., Drozdovsky, I.~O., Turnshek, D.~A., 
Nestor, D.~B., \& Pettini, M.\ 2004, ApJ, 600, 613 

\bibitem{} Shapley, A.~E., Erb, D.~K., Pettini, M., Steidel, C.~C., 
\& Adelberger, K.~L. 2004, ApJ, 612, 108

\bibitem{} Shapley, A.~E., Steidel, C.~C., Erb, D.~K., Reddy, N.~A.,
Adelberger, K.~L., Pettini, M., Barmby, P., \& Huang, J. 2005, 626, 698

\bibitem{} Simcoe, R.~A., Sargent, W.~L.~W., \& Rauch, M. 2004, ApJ, 606, 92

\bibitem{} Songaila, A. 2006, AJ, 131, 24

\bibitem{} Stasi{\' n}ska, G.\ 2005, A\&A, 434, 507 

\bibitem{} Steidel, C.~C., Adelberger, K.~L., Giavalisco, M., 
Dickinson, M., \& Pettini, M. 1999, ApJ, 519, 1

\bibitem{} Steidel, C.~C., Adelberger, K.~L., Shapley, A.~E., 
Erb, D.~K., Reddy, N.~A., \& Pettini, M. 2005, ApJ, 626, 44

\bibitem{} Steidel, C.~C., Giavalisco, M., Pettini, M., Dickinson, M.,
\& Adelberger, K.~L. 1996, ApJ, 462, L17

\bibitem{} Steidel, C.~C., Shapley, A.~E., Pettini, M.,  Adelberger, K.~L.,
Erb, D.~K., Reddy, N.~A., \& Hunt, M.~P. 2004, ApJ, 604, 534

\bibitem{}  Swinbank, A.~M., 
Smail, I., Chapman, S.~C., Blain, A.~W., Ivison, R.~J., \& Keel, W.~C.\ 
2004, ApJ, 617, 64 

\bibitem{} Wild, V., Hewett, P.~C., \& Pettini, M. 2006, MNRAS, in press (astro-ph/0512042)

\bibitem{} Wolfe, A.~M.,  Gawiser, E., \& Prochaska, J.~X. 2003, ApJ, 593, 235

\bibitem{} Wolfe, A.~M., Prochaska, J.~X., \& Gawiser, E. 2005, ARAA, 43, 861
 
\bibitem{} Wolfe, A.~M., Turnshek, D.~A., Smith, H.~E., \& Cohen, R.~D. 1986,
ApJS, 61, 249

\bibitem{} Wyse, R.~F.~G., \& Gilmore, G. 1995, AJ, 110, 2771

\bibitem{} Zwaan, M.~A., van der Hulst, J.~M., Briggs, F.~H., Verheijen, M.~A., W., \& Ryan-Weber, E.~V.  2005, MNRAS, 364, 1467

}
\end{thebibliography}
\end{document}